\documentclass[
    aps,prb,10pt,
    twocolumn,reprint,
superscriptaddress,floatfix,amsmath,amssymb
]{revtex4-2}

\usepackage{graphicx}\graphicspath{{figures/}}
\usepackage[caption=false]{subfig}

\usepackage{dcolumn}\usepackage{bm}\usepackage{hyperref}\usepackage{mhchem}\usepackage{color}
\usepackage{siunitx}
\usepackage{algorithm}
\usepackage{algpseudocode}
\usepackage[export]{adjustbox}

\newcommand{\Tr}{\operatorname{Tr}}\newcommand{\Hhat}{\hat{H}}\newcommand{\Vhat}{\hat{V}}\newcommand{\Dcal}{\mathcal{D}}\newcommand{\Tcal}{\mathcal{T}}\newcommand{\drm}{\mathrm{d}}
\newcommand{\iwn}{\mathrm{i}\omega_{n}}

\begin{document}

\title{Interaction expansion inchworm Monte Carlo solver for lattice and impurity models}

\def\umphys{Department of Physics, University of Michigan,
    Ann Arbor, MI 48109, USA
}\def\telaviv{School of Chemistry, Tel Aviv University,
    Tel Aviv 69978, Israel
}

\author{Jia Li}
\affiliation{\umphys}

\author{Yang Yu}
\affiliation{\umphys}

\author{Emanuel Gull}
\affiliation{\umphys}

\author{Guy Cohen}
\affiliation{\telaviv}

\date{\today}

\begin{abstract}
Multi-orbital quantum impurity models with general interaction and hybridization terms appear in a wide range of applications including embedding, quantum transport, and nanoscience.
However, most quantum impurity solvers are restricted to a few impurity orbitals, discretized baths, diagonal hybridizations, or density--density interactions.
Here, we generalize the inchworm quantum Monte Carlo method to the interaction expansion and explore its application to typical single- and multi-orbital problems encountered in investigations of impurity and lattice models.
Our implementation generically outperforms bare and bold-line quantum Monte Carlo algorithms in the interaction expansion.
    So far, for the systems studied here, it remains inferior to the more specialized hybridization expansion and auxiliary field algorithms.
    The problem of convergence to unphysical fixed points, which hampers so-called bold-line methods, is not encountered in inchworm Monte Carlo.
\end{abstract}

\maketitle

\section{Introduction}
The efficient solution of quantum impurity problems is one of the fundamental challenges in computational condensed matter physics.
Impurity models, in general terms, consist of a `local' Hamiltonian with a small number of interacting orbitals, coupled via a hybridization term to an infinite number of noninteracting reservoir orbitals.
Much of the current interest in impurity models stems from their use in embedding theories \cite{Georges96,Kotliar06,Zgid17}.
They are also used directly, in the description of, e.g., atoms adsorbed onto surfaces \cite{Brako81,Langreth91} and magnetic impurities embedded in a metallic host \cite{Anderson61}.
Their nonequilibrium properties are important in nanoscience and quantum transport, where impurity models are used to describe quantum dots \cite{datta_electronic_1997,Hanson07} and molecular conductors \cite{aviram_molecular_1974,nitzan_electron_2003,cohen_greens_2020}.

`Diagrammatic' Monte Carlo (DiagMC) methods, i.e. Monte Carlo methods that sample diagrammatic perturbation theories \cite{Prokofev98A,Prokofev98B}, have proven to be effective solution methods for quantum impurity models, along with exact diagonalization (ED) \cite{Caffarel94,Iskakov18}, renormalization group \cite{Bulla08,Bulla99}, tensor network \cite{Wolf14,Bauernfeind17,Feiguin04}, and several quantum chemistry approaches \cite{Zgid12,Zhu19,Shee19,Shee21}.
In particular, continuous-time methods (CT-QMC) \cite{Gull11_RMP} are currently the gold standard for the solution of impurity problems generated by most embedding approaches.
This includes the interaction expansion (CT-INT) \cite{Rubtsov04,Rubtsov05}, the hybridization expansion (CT-HYB) \cite{Werner06A,Werner06B,Haule07}, and the auxiliary field method (CT-AUX) \cite{Gull08,Gull11_submatrix}, each of which has different advantages and regimes of applicability \cite{Gull11_RMP}.

These methods nevertheless reach their limit in impurity models with general interactions and off-diagonal hybridizations, as they appear in \textit{ab initio} embedding setups.
In these systems, the various expansions generically encounter sign problems.
This means the computational cost of simulations increases exponentially as a function of system size \cite{Gull11_RMP}, interaction strength, inverse temperature, or some other control parameter.
Diagrammatic Monte Carlo methods based on expansions of an observable such as the Green's function \cite{VanHoucke2010,Rossi2017b,Li20}, the self-energy \cite{Moutenet2018,Simkovic2019,Rossi2018}; or bold-line strategies \cite{Prokofev2008b,Gull10,Cohen2014a,Cohen2014b} may then present an alternative path towards the solution.
These methods incorporate resummation techniques into their design, and therefore truncation of the perturbation series at relatively low orders becomes accurate.
The cost is typically giving up on absolute convergence\cite{Prokofev2008b} and relying on self-consistency conditions that may converge to an unphysical fixed point \cite{Kozik2015}.
Many of these techniques also face problems with series convergence, so that their use in practice requires analytical continuation \cite{Rossi18B}.

The inchworm Monte Carlo method \cite{Cohen2015,Eidelstein2019} is a kind of resummation technique, much like bold-line methods.
However, in contrast to bold-line methods, inchworm methods lack a self-consistency cycle and cannot converge to unphysical fixed points.
As shown in ref.~\cite{Eidelstein2019}, the inchworm hybridization expansion is able to address multiorbital systems where CT-HYB suffers from a severe sign problem.

In this paper, we present an inchworm method constructed around the \emph{interaction} expansion.
We examine the performance of the method in comparison to bare and bold-line interaction expansion impurity solvers, as well as the CT-HYB \cite{Werner06A,Werner06B} and CT-AUX \cite{Gull08} continuous-time methods.
We find that the method performs better than the diagrammatic Monte Carlo  methods, but that for the simple models investigated here CT-AUX and CT-HYB outperform the inchworm method.
Nevertheless, the flexibility of the framework presented here is such that many generalizations and improvements are possible, including combination with some of the approaches mentioned above.
It may therefore be the first step on a path to the development of highly efficient new methods in the future.

The paper will proceed as follows.
Sec.~\ref{sec:method} explains the main idea of inchworm methods, and shows how it can be used in the context of the interaction expansion.
Section~\ref{sec:results} presents applications to impurity models, and Sec.~\ref{sec:conclusions} discusses conclusions.

\section{Method}\label{sec:method}

\begin{figure}[tb]
    \includegraphics[width=0.8\linewidth,trim=0 0 0 1.55cm, clip]{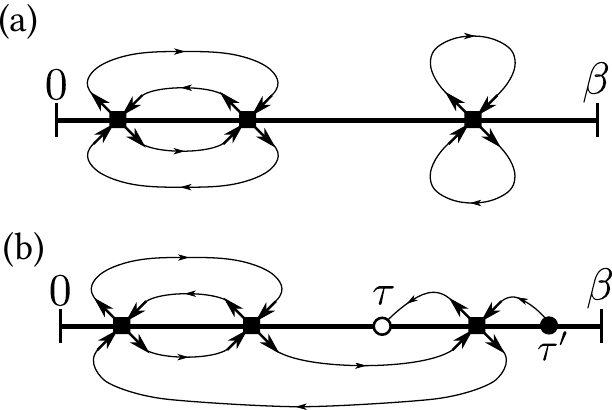}
    \caption{Examples of bare Feynman diagrams.
        (a): disconnected third order diagram for $Z$.
        (b): (connected) third order diagram for
        $G(\tau,\tau')$.
        Filled squares are vertices representing
        $U_{ijkl}$, lines with arrows bare propagators representing $G_0$, and
        open/closed circles represent external operators.
    }
    \label{fig:diagrams}
\end{figure}

\subsection{Imaginary time perturbation theory}
We derive our method for the generic electronic structure Hamiltonian
\begin{equation}
    \Hhat  = \Hhat_0 + \Vhat, \label{eq:hamiltonian}
\end{equation}
with
\begin{align}
    \quad\Hhat_0 & = \sum_{ij,\sigma} h_{ij} \hat{c}^\dagger_{i\sigma} \hat{c}_{j\sigma},                                                                           \\
    \quad\Vhat   & = \frac{1}{2}\sum_{ijkl}\sum_{\sigma\sigma'} U_{ijkl} \hat{c}^\dagger_{i\sigma} \hat{c}^\dagger_{k\sigma'} \hat{c}_{l\sigma'} \hat{c}_{j\sigma}.
\end{align}
Here, $\hat{c}_i^{\dagger},\hat {c}_i$ are electron creation and annihilation operators in orbital $i$, $h$ is the single-particle Hamiltonian, and $U$ the electronic interaction tensor.
In the context of lattice models and electronic structure setups, $U$ spans all orbitals.
For impurity models, $U$ is restricted to a small `impurity' subspace.

We take a perturbative approach following the interaction expansion formalism,
by treating the noninteracting Hamiltonian $\Hhat_0$ as the unperturbed system and the interaction $\Vhat$ as the perturbation.
The partition function of the system at inverse temperature $\beta$ can be expanded as a
series in the interaction picture~\cite{Abrikosov1965,Mahan2000},
\begin{align}
     Z  &= \Tr e^{-\beta\Hhat} = \Tr [e^{-\beta\Hhat_0} \hat U_I(\beta)] = Z_0 \langle \hat{U}_I (\beta) \rangle_0, \label{eq:Zfull-expansion}
    \\
    \begin{split}
         \hat{U}_I(\beta)  &:= e^{\beta \Hhat_0} e^{-\beta\Hhat} = \sum_{k=0}^{\infty}\frac{(-1)^{k}}{k!}\int_{0}^{\beta}\drm\tau_{1}\int_{0}^{\beta}\drm\tau_{2} \\
        &\cdots\int_{0}^{\beta}\drm\tau_{k}                                                                                                                       \Tcal_{\tau}\{\hat{V}_{I}(\tau_{1})\hat{V}_{I}(\tau_{2})\cdots\hat{V}_{I}(\tau_{k})\}, \end{split}
\end{align}
where the subscript $I$ denotes operators in the interaction picture,
$\hat U_I(\tau) = e^{\tau \Hhat_0} e^{-\tau\Hhat}$ the time evolution operator,
$Z_0 = \Tr e^{-\beta \Hhat_0}$ the noninteracting partition function,
$\langle\cdot\rangle_0 = Z_0^{-1} \Tr [e^{-\beta\Hhat_0}(\cdot)]$ the
noninteracting thermal expectation value, and $\Tcal_{\tau}$ the time
ordering operator.
Similarly, the electronic Green's function
in imaginary time, defined as \begin{equation}
    G_{ij}(\tau,\tau')=G_{ij}(\tau-\tau')=-\langle \Tcal_{\tau}\hat{c}_i(\tau)\hat{c}^{\dagger}_j(\tau'+0^{+})\rangle
    \label{eq:gf-def}
\end{equation}
where $\langle\cdot\rangle = Z^{-1}  \Tr[e^{-\beta\Hhat}(\cdot)]$, can be expanded as~\cite{Mahan2000,Rubtsov05,Gull11_RMP}
\begin{equation}
    \begin{aligned}
         G(\tau,\tau') &=-\frac{Z_{0}}{Z}\sum_{k=0}^{\infty}\frac{(-1)^{k}}{k!}\int_{0}^{\beta}\drm\tau_{1}\int_{0}^{\beta}\drm\tau_{2}\cdots\int_{0}^{\beta}\drm\tau_{k}\\
         & \times\langle\Tcal_{\tau}\hat{c}_{I}(\tau)\hat{c}_{I}^{\dagger}(\tau')\hat{V}_{I}(\tau_{1})\hat{V}_{I}(\tau_{2})\cdots\hat{V}_{I}(\tau_{k})\rangle_{0}.
    \end{aligned}\label{eq:Gfull-expansion}
\end{equation}

It will be convenient to introduce a parameter $\theta \in [0,\beta]$ and define an
auxiliary partition function
\begin{equation}
    Z_\theta := Z_0\langle \hat{U}_I(\theta) \rangle_0 = \Tr[e^{-(\beta-\theta)\Hhat_0} e^{-\theta\Hhat}].
    \label{eq:Zaux-def}
\end{equation}
Since $\hat{U}_I(0)$ is the identity operator,
$Z_\theta$ connects $Z_0=Z_0\langle \hat{U}_I(0) \rangle_0$ and
$Z=Z_0\langle \hat{U}_I(\beta) \rangle_0$ continuously via the parameter
$\theta$, such that $Z_{\theta= 0} = Z_0$, $Z_{\theta=\beta} = Z$.
With $\hat{U}_I$ from Eq.~(\ref{eq:Zfull-expansion}),
$Z_\theta$ can be expanded as
\begin{equation}
    \begin{aligned}
        Z_\theta &= Z_{0}\sum_{k=0}^{\infty}\frac{(-1)^{k}}{k!}\int_{0}^{\theta}\drm\tau_{1}\int_{0}^{\theta}\drm\tau_{2}
\cdots\int_{0}^{\theta}\drm\tau_{k}\\
        &\times
        \langle\Tcal_{\tau}\hat{V}_{I}(\tau_{1})\hat{V}_{I}(\tau_{2})\cdots\hat{V}_{I}(\tau_{k})\rangle_{0},
    \end{aligned}\label{eq:Zaux-expansion}
\end{equation}
corresponding to the expression for $Z$ with all upper integration bounds replaced by
$\theta$.
$Z_\theta=\Tr[e^{-(\beta-\theta)\Hhat_0} e^{-\theta\Hhat}]$ can be understood as a
trace of a `partially dressed' time
evolution: from 0 to $\theta$ the system is propagated with
the full Hamiltonian $\Hhat$, and then from $\theta$ to $\beta$ with the
noninteracting Hamiltonian $\Hhat_0$.

The equivalent change of the integration bounds in Eq.~(\ref{eq:Gfull-expansion}) to $\theta$ defines an
auxiliary Green's function
\begin{equation}
    \begin{aligned}
         G_\theta(\tau,\tau') &=-\frac{Z_{0}}{Z_\theta}\sum_{k=0}^{\infty}\frac{(-1)^{k}}{k!}\int_{0}^{\theta}\drm\tau_{1}\int_{0}^{\theta}\drm\tau_{2}\cdots\int_{0}^{\theta}\drm\tau_{k} \\
        &\times\langle\Tcal_{\tau}\hat{c}_{I}(\tau)\hat{c}_{I}^{\dagger}(\tau')\hat{V}_{I}(\tau_{1})\hat{V}_{I}(\tau_{2})\cdots\hat{V}_{I}(\tau_{k})\rangle_{0}.
    \end{aligned}\label{eq:Gaux-expansion}
\end{equation}
$G_\theta$ continuously connects the noninteracting Green's
function at $\theta=0$ to the full
Green's function $G$ at $\theta=\beta$.
In Appendix~\ref{app:Gaux-explicit} we show an explicit non-perturbative
definition of $G_\theta$.
Since $\theta$ breaks time-translational
invariance, $G_\theta$ is defined as a function of two time parameters and cannot be defined as function of a single time parameter
as in Eq.~(\ref{eq:gf-def}).

\subsection{Diagrammatic evaluation of auxiliary quantities\label{subsec:diagram_rules}}
The expansions of physical quantities $Z$ and $G$, when applied to
the electronic Hamiltonian (\ref{eq:hamiltonian}), can be
represented graphically as a sum over
Feynman diagrams in the usual way~\cite{Abrikosov1965}.
A diagram at order $k$ is composed of $k$ interaction vertices representing $U_{ijkl}$,
each assigned to an imaginary time index $\tau_i\in[0,\beta]$, $i=1,\ldots,k$.
Propagator lines representing the noninteracting Green's function $G_0$
connect these vertices.
For the partition function $Z$, the noninteracting expectation values in
Eq.~(\ref{eq:Zfull-expansion}) can be evaluated using Wick's theorem, which
generates closed `vacuum' diagrams which can be either connected or
disconnected.
The Green's function expansion in Eq.~(\ref{eq:Gfull-expansion}) involves two
`external' operators $\hat{c}_I(\tau)$ and $\hat{c}^\dagger_I(\tau')$ which
become external `legs' in Feynman diagrams, and the disconnected components
are canceled by the partition function diagrams of $Z$ in the denominator,
leaving diagrams in which all internal vertices and external legs are fully
connected~\cite{Abrikosov1965,Negele1988}.
Figure~\ref{fig:diagrams} shows examples of such `bare' Feynman diagrams.

\begin{figure}
    \includegraphics[width=0.5\columnwidth]{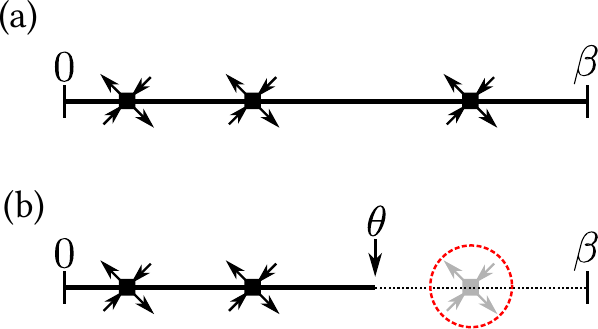}
    \caption{Illustration of valid vertex coordinates on the imaginary time axes.
        (a): Vertices can take any $\tau$ value from $0$ to $\beta$ in the diagrammatic expansion for $G$ or $Z$;
        (b): Vertices can only take $\tau$ values from $0$ to $\theta$ in the diagrammatic expansion for $G_\theta$ or $Z_\theta$.
        Dashed red circle indicates an invalid vertex.}
    \label{fig:vertices}
\end{figure}

Since expansions of the auxiliary quantities, Eqs.~(\ref{eq:Zaux-expansion}) and
(\ref{eq:Gaux-expansion}), only differ from Eqs.~(\ref{eq:Zfull-expansion}) and
(\ref{eq:Gfull-expansion}) in the integration bounds of internal time indices,
the same diagram rules can be applied to compute $Z_\theta$ and $G_\theta$, as
long as the vertices $U$ are confined to the imaginary time interval
$[0,\theta]$, as illustrated in Fig.~\ref{fig:vertices}.
The expansions of $G$ and $G_\theta$ can be formally written as
\begin{equation}
    \begin{aligned}
        G(\tau,\tau') &= \sum_{k=0}^{\infty} \frac{(-1)^k}{k!} \int_0^{\beta}\drm\tau_1 \int_0^{\beta}\drm\tau_2 \cdots \int_0^{\beta}\drm\tau_k \\
        &\times D^{\mathrm{bare}}(\tau,\tau';\tau_1, \tau_2, \ldots, \tau_k),\\
        G_{\theta}(\tau,\tau') &= \sum_{k=0}^{\infty} \frac{(-1)^k}{k!} \int_0^{\theta}\drm\tau_1 \int_0^{\theta}\drm\tau_2 \cdots \int_0^{\theta}\drm\tau_k \\
        &\times D^{\mathrm{bare}}(\tau,\tau';\tau_1, \tau_2, \ldots, \tau_k),
    \end{aligned}\label{eq:formal-bare-expansion}
\end{equation}
where $D^\mathrm{bare}$ denotes the sum of all connected bare
diagrams~\cite{Abrikosov1965,Negele1988}.

\begin{figure*}[bth]
    \includegraphics[width=\linewidth]{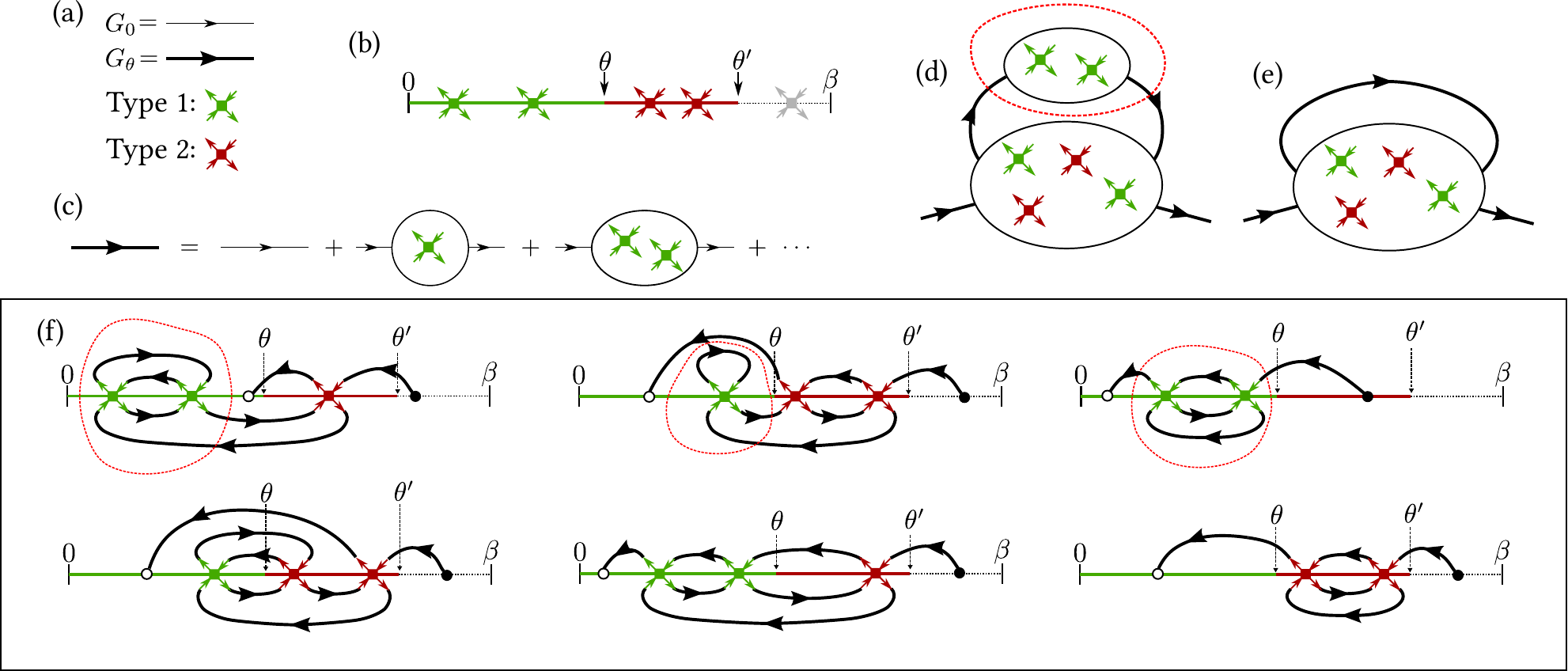}
    \caption{Diagram rules for the inchworm expansion
        from $G_\theta$ to
        $G_{\theta'}$ with $\theta < \theta'$. (a) Thin lines
        stand for the bare propagator $G_0$, and the
        `dressed' lines for $G_\theta$. Green (red) crosses represent Type 1 (2)
        vertices.
        (b) Type 1 vertices can only be inserted in $[0, \theta]$ (green segment),
        and Type 2 vertices in $(\theta,\theta']$ (red segment). Neither type of
        vertices are allowed in the dashed segment.
        (c) Each `dressed' line can be expanded into a bare series following
        Eq.~(\ref{eq:Gaux-expansion}), in terms of connected diagrams with only Type
        1 vertices.
        Diagram (d) is an example of a connected diagram that needs to be excluded
        from the inchworm expansion, since it is already included in Diagram (e).
        In Box (f), the top row of diagrams are excluded by the diagram rules, where
        the overcounted components are enclosed by dashed red curves; the bottom
        row shows valid diagrams in the expansion
        Eq.~(\ref{eq:formal-inch-expansion}).
    }
    \label{fig:diagram-rules-inch}
\end{figure*}

Inchworm diagrammatics aims to reuse knowledge of propagators up to some time in order to calculate propagators to a larger time \cite{Cohen2015}.
In the context of the interaction expansion, we assume knowledge of $G_\theta$ for some time $\theta$, and aim to express $G_{\theta'}$ for $\theta'>\theta$.
Crucially, we write the diagrammatic series for $G_{\theta'}$ in terms of $G_\theta$ rather than $G_0$.
The series is therefore partially `dressed', and every contribution to $G_\theta$ contains infinitely many bare diagram components.
The latter diagrams are valid terms in the standard bare series for $G_\theta$, in which all internal vertices residing in the interval $[0,\theta]$ are already accounted for at order zero.
As panels (d) and (e) in Fig.~\ref{fig:diagram-rules-inch} demonstrate, some diagram topologies would be overcounted if the unmodified diagram rules were applied to this expansion.
The diagrammatics therefore needs to be modified.

We now summarize the diagram rules for computing $G_{\theta'}$ from $G_\theta$.
For a given set of vertices $U$ at
$\tau_1,\ldots,\tau_k\in[0,\theta']$ and external operators
$\hat{c}$, $\hat{c}^\dagger$ at $\tau$, $\tau'$:
\begin{enumerate}
    \item Generate all possible graphs by connecting vertices and operators with propagator lines.
    \item Eliminate all disconnected graphs.
    \item Sort the vertices into two categories:
          \begin{itemize}
              \item `Type 1' if $0 < \tau_i < \theta$,
              \item `Type 2' if $\theta < \tau_i < \theta'$.
          \end{itemize}
    \item Eliminate all graphs that only contain Type-1 vertices.
    \item Eliminate all graphs that contain subgraphs of Type-1 vertices connected with exactly two propagators to the remainder of the graph.
\end{enumerate}
Figure~\ref{fig:diagram-rules-inch} illustrates these rules.
The first three rules are equivalent to those of bare perturbation theory~\cite{Abrikosov1965,Negele1988}, and the
additional rules exclude overcounted diagrams.
Note that rule 5 is analogous to the `skeleton' diagram rules of the
self-energy for bold-line perturbation theory~\cite{Luttinger1960}.

The diagrammatic series can be formally written as
\begin{equation}
    \begin{aligned}
        G_{\theta'}(\tau,\tau') & = \sum_{k=0}^{\infty} \frac{(-1)^k}{k!} \int_0^{\theta'}\drm\tau_1 \int_0^{\theta'}\drm\tau_2 \cdots \int_0^{\theta'}\drm\tau_k \\
                                & \times D_{\theta}(\tau,\tau';\tau_1, \tau_2, \ldots, \tau_k),
    \end{aligned}
    \label{eq:formal-inch-expansion}
\end{equation}
where $D_{\theta}$ denotes the sum of all diagrams following the updated diagram rules in which $G_\theta$ is used as the propagator.
We emphasize here that all the internal time indices $\tau_1,\ldots,\tau_k$ are integrated from 0 to $\theta'$, whereas the external indices $\tau$ and $\tau'$ are unconstrained and can take on all values between 0 and $\beta$.

For $\theta'\to\theta$, $G_{\theta'}$ continuously approaches $G_\theta$, and the expansion Eq.~\eqref{eq:formal-inch-expansion} includes substantially fewer diagrams than the bare expansion Eq.~\eqref{eq:formal-bare-expansion}.
As we will show in Sec.~\ref{sec:results}, $G_\theta$ is typically a much better starting point for a perturbation expansion of $G_{\theta'}$ than $G_0$; and this is especially true when $\theta'-\theta\ll\beta$.

\subsection{Inchworm Monte Carlo algorithm}

The ability to efficiently obtain $G_{\theta'}$ from $G_\theta$ with $\theta<\theta'$ suggests an iterative algorithm where $N$ simulations are performed sequentially with different values of $\theta$: $\theta_1=0<\theta_2<\ldots<\theta_N=\beta$.
For each $N>0$, $G_{\theta_n}$ is obtained from $G_{\theta_{n-1}}$.
We expect the similarity between $G_{\theta_n}$ and $G_{\theta_{n-1}}$ to reduce the number of diagrams that needs to be evaluated, thereby reducing the difficulty of the simulation.
In analogy to the inchworm algorithm for the hybridization expansion~\cite{Cohen2015,Eidelstein2019},
which utilizes the same strategy for gradually increasing the range of propagators, we call
the parameter $\theta$ the `inchworm time', and refer to the expansion from Eq.~(\ref{eq:formal-inch-expansion}) as the `inchworm expansion'.
For any choice of intermediate time points the final solution is guaranteed to be exact if
(1) the perturbation series converges in each inchworm expansion calculation, and (2) the series is
computed to all orders.

By making the difference in inchworm time $\Delta \theta=\theta' - \theta$ sufficiently small, such that $G_{\theta'}$ is well approximated by $G_{\theta}$, we observe that in practice the first assumption is satisfied for all systems we study in Sec.~\ref{sec:results}.
In Appendix~\ref{app:skeleton}, we connect the convergence of inchworm series to the skeleton expansion~\cite{Luttinger1960}. Unlike inchworm results, which are obtained by a simple forward propagation, skeleton series are typically obtained self-consistently and may converge to an unphysical fixed point~\cite{Kozik2015}.

The summation of  diagrams to all orders is not feasible for most systems of interest. However, one may  calculate the contribution to an observable of interest as a function of expansion order. If a decay of the contribution is observed as a function of expansion order, results can be obtained without summing all diagrams to infinite order. Section~\ref{sec:results} shows examples where this procedure succeeds, and systems where contributions do not decay within the
accessible orders.

We briefly comment on the choice of inchworm times. With a uniform discretization, $\beta/\Delta\theta)$ Monte Carlo simulations are needed for the final result.
As evident from Eq.~\ref{eq:formal-inch-expansion}, the inchworm expansion stays exact for any choice of $\Delta \theta$ for converged series.
This is a major difference from certain Monte Carlo algorithms that employ a Trotter--Suzuki decomposition, where the time discretization $\Delta\theta$ introduces an approximation and final results need to be extrapolated to the limit of $\Delta\theta\rightarrow 0$ (e.g. Ref. \cite{BSS81}).
The choice of time grid is therefore given by the following empirical considerations: If $\Delta \theta$ is chosen small, the expansion becomes efficient but more simulations are needed for the final result.
On the other hand, if $\Delta \theta$ is large, higher diagram order is required to obtain the same quality of results.
In practice, in the simulations discussed in Section~\ref{sec:results} we often chose $8-16$ time slices, far fewer than in typical Trotter--Suzuki simulations.

\begin{figure}
    \includegraphics[width=0.9\linewidth]{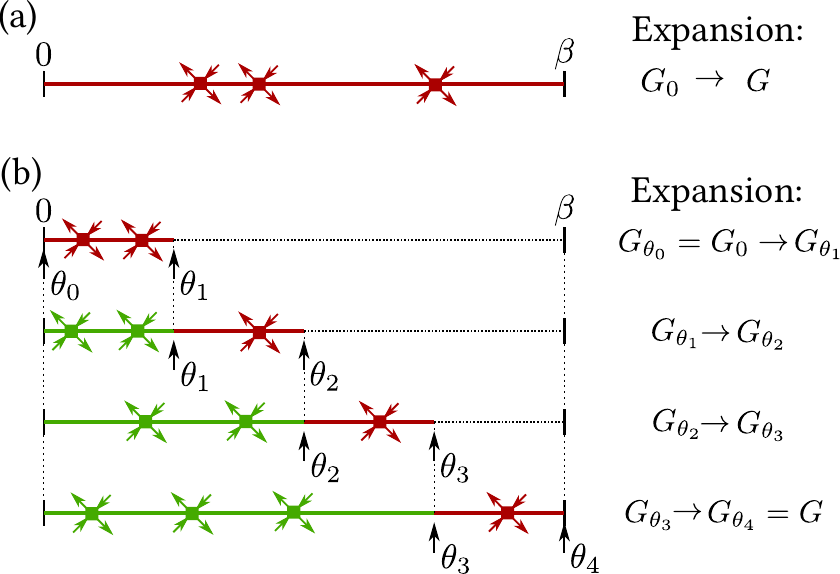}
    \caption{Schematic comparison of bare and inchworm Monte Carlo for Green's function.
        (a) shows a vertex configuration for the bare expansion in
        Eq.~\ref{eq:Gfull-expansion} which is equivalent to an inchworm simulation
        with $N=1$.
        (b) shows configurations for an inchworm
        Monte Carlo simulation with $N=4$ at each inchworm step.
        In Monte Carlo samplings of each expansion, Type-1 (Type-2) vertices are sampled
        in green (red) segments on the imaginary time axis.
    }
    \label{fig:inching}
\end{figure}

A complete inchworm simulation proceeds as follows.
We first construct two imaginary time grids: one `inchworm grid' $\{\theta_n| n = 0, \ldots, N; \theta_{n} > \theta_{n-1} \} $ for the sequence of inchworm times $\theta$, and one `interpolation grid' $\{\tau_i| i = 0, \ldots, N_\tau\}$ for measuring and interpolating the auxiliary Green's function.
The final Green's function is then computed via $N$ `inchworm steps':
In the $n$-th step, we perform an inchworm expansion of $G_{\theta_n}$ with respect to $G_{\theta_{n-1}}$, and calculate $G_{\theta_n}(\tau_i, \tau_j)$ for each pair of $i,j=0,\ldots,N_\tau$ using Monte Carlo as detailed in Sec.~\ref{subsec:MC}, with the noninteracting initial condition $G_{\theta_0} = G_0$.
Figure~\ref{fig:inching} illustrates the `inching' process, in comparison with the bare expansion which is equivalent to performing only a single inchworm step.
$G_{\theta_n}$ is evaluated on the interpolation grid for continuous-time evaluations within the next inchworm step.

For simplicity, we chose equidistant time points for both grids, and perform linear interpolation for measured auxiliary Green's functions.
This provides decent accuracy at high temperatures.
Since $G_{\theta}(\tau,\tau')$ is generally not smooth when $\tau=\theta$ or $\tau'=\theta$, we required the interpolation grid $\{\tau_i\}$ to include all points on the inchworm grid $\{\theta_n\}$ so that the sharp corners at these points are well-resolved.
Nevertheless, we note that nothing in the algorithm precludes using a nonuniform (e.g. Chebyshev) interpolation grid \cite{boehnke2011,shinaoka2017,gull2018}, and this will most likely be advantageous at lower temperatures.

\subsection{Continuous-time Monte Carlo evaluation of inchworm expansions}\label{subsec:MC}

We evaluate each inchworm expansion step (\ref{eq:formal-inch-expansion}) in a standard diagrammatic/continuous-time quantum Monte Carlo approach~\cite{Prokofev98A,Prokofev98B,Rubtsov04,Rubtsov05,Gull08,Gull11_RMP,Rossi2017b}.
We employ a finite cutoff $k_\mathrm{max}$ of the expansion order, and perform Monte Carlo importance sampling of the internal spacetime coordinates following the \textit{a priori} distribution
\begin{equation}
    p(\mathcal{C}) \propto |D_{\theta}(\tau,\tau';\mathcal{C})|.
\end{equation}
Here, $\mathcal{C} = \{\tau_1,\ldots,\tau_k\}$ is a Monte Carlo configuration.
Since $D_{\theta}$ has varying signs due to its fermionic nature, the absolute value is necessary to ensure $p(\mathcal{C})$ is positive, whereas the fermionic sign $\operatorname{sgn} (D_{\theta})$ enters measurements of all physical observables.
For a given vertex configuration, $D_{\theta}$ is computed explicitly by summing over all proper inchworm diagrams according to the diagram rules of Subsec.~\ref{subsec:diagram_rules}.
In our implementation, we rely on a graph theory library to precompute and save all valid diagram topologies for each expansion order.
We generate Monte Carlo samples as a Markov chain using the Metropolis--Hastings algorithm.
From each configuration $\mathcal{C}$, a new configuration $\mathcal{C}'$ is proposed following some proposal probability distribution $w^\mathrm{prop}(\mathcal{C}'|\mathcal{C})$.
To ensure detailed balance, an acceptance ratio $R$ is calculated after each proposal as
\begin{equation}
    R(\mathcal{C}'|\mathcal{C}) = \frac{w^\mathrm{prop}(\mathcal{C}|\mathcal{C}')p(\mathcal{C}')}{w^\mathrm{prop}(\mathcal{C}'|\mathcal{C})p(\mathcal{C})}.
\end{equation}
The proposal $\mathcal{C}\to\mathcal{C}'$ is accepted with probability
\begin{equation}
    w^\mathrm{acc}(\mathcal{C}'|\mathcal{C}) = \min(1, R(\mathcal{C}'|\mathcal{C})).
\end{equation}
This ensures the detailed balance of the Markov process, i.e.
\begin{equation}
    w(\mathcal{C}'|\mathcal{C})p(\mathcal{C})=w(\mathcal{C}|\mathcal{C}')p(\mathcal{C}'),
\end{equation}
where
\begin{equation}
    w(\mathcal{C}'|\mathcal{C}) = w^\mathrm{acc}(\mathcal{C}'|\mathcal{C})w^\mathrm{prop}(\mathcal{C}'|\mathcal{C}).
\end{equation}
This procedure generates samples drawn from the equilibrium distribution $p(\mathcal{C})$.
We employ the same Monte Carlo updates as in CT-INT~\cite{Rubtsov05}, which guarantee ergodicity for all the systems studied in this work.
Those include random insertions and removals of a single vertex or a pair of vertices.
The auxiliary Green's function $G_\theta$ is measured during the Monte Carlo procedure and normalized against quantities that are analytically tractable, such as low-order diagrams.

\section{Results}\label{sec:results}

For most of the discussion below, we limit ourselves to small isolated lattices such as the Hubbard atom, dimer, and trimer: i.e., models with one to three spin-half orbitals.
These are systems that, in the case of the interaction expansion inchworm method, have the same complexity as quantum impurity models with the same number of orbitals.
However, unlike impurity models, which also feature an infinite noninteracting bath, they can be exactly diagonalized without further approximations, such that a reliable benchmark is available.
In all cases, the hopping parameter $t$ is used as the unit of energy.

\begin{figure*}[tbh]
    \includegraphics[width=0.9\linewidth]{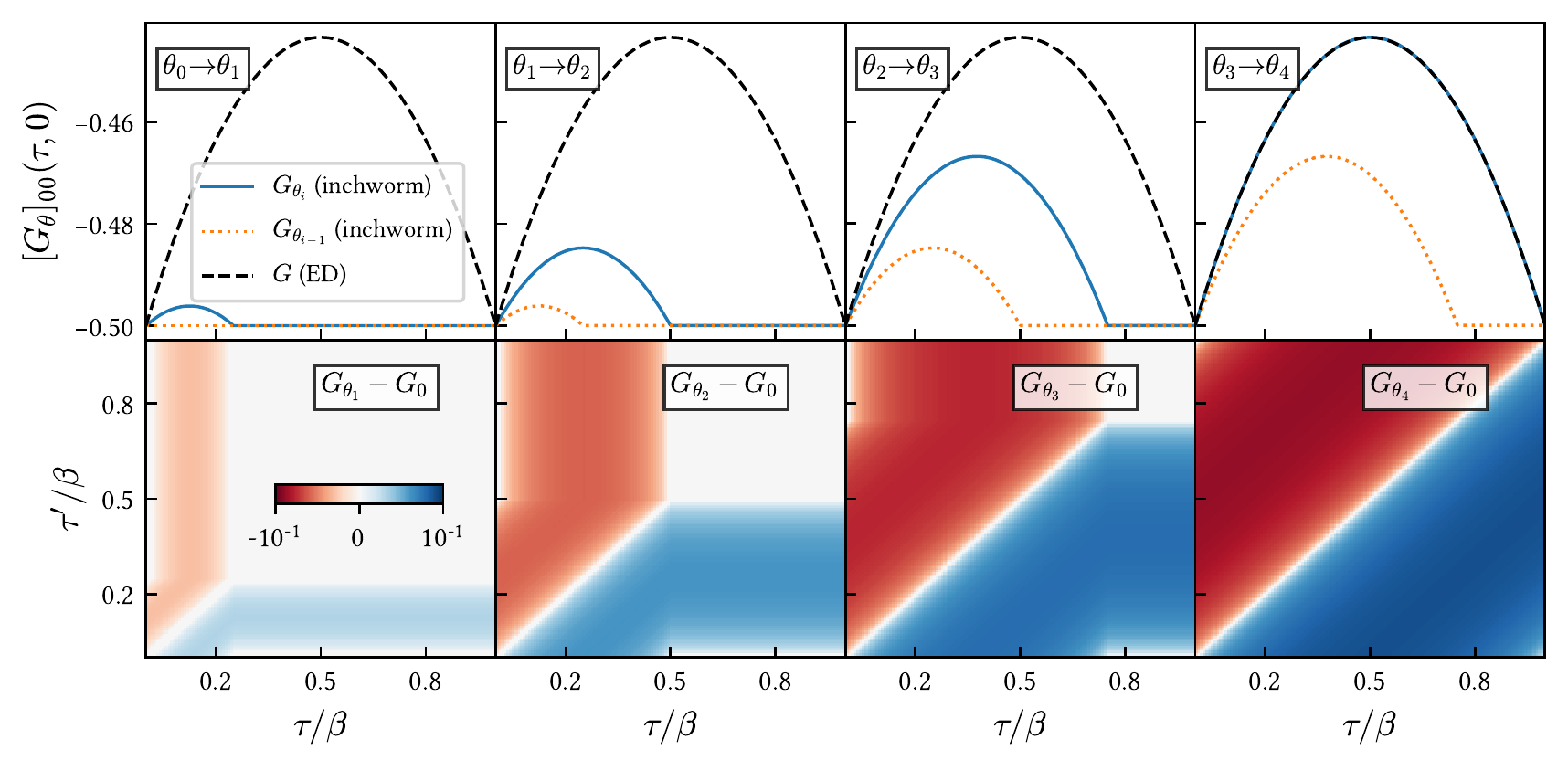}
    \caption{Auxiliary Green's functions of a Hubbard atom at half-filling, $U=1$, $\beta=2$.
        Top panel: $G_{\theta_{i}}(\tau,0)$ at each step, compared to the previous inchworm step and
        the exact $G$.
        Bottom panel: $G_{\theta_{i}}(\tau,\tau')-G_0(\tau,\tau')$ in both $\tau$ and $\tau'$ dimensions.
        \label{fig:atomsteps}}
\end{figure*}
We start our discussion of the results with Fig.~\ref{fig:atomsteps}, which illustrates the two-time Green's function $G_\theta$ of Eq.~\ref{eq:formal-bare-expansion}.
Results are shown for the half-filled Hubbard atom (the one-site Hubbard model) at  interaction $U=1$ and inverse temperature $\beta=2$, for a discretization of $\Delta \theta =\beta/4$.
Intermediate inchworm steps are shown on the left three panels, and the final inchworm step for the last one.
Black dashed lines in the upper panels show the exact result obtained from ED.
$G_{\theta_{i-1}}(\tau,0)$ from the previous step is plotted in red dotted lines, and blue lines show the currently computed result $G_{\theta_{i}}(\tau,0)$.
The bottom panels illustrate the two-time function $G_\theta$ as a contour plot with the two time arguments.

The inchworm algorithm starts from the noninteracting Green's function and, while `inching' forward, gradually advances towards the interacting Green's function.
The propagation breaks the time-translational symmetry, such that only the initial and the final solution are time-translation invariant: i.e., $G_{\theta_{4}}(\tau,\tau')=G_{\theta_{4}}(\tau-\tau')$, and the same is true for $G_0$; but this is not the case for $G_{\theta_{1}}$, $G_{\theta_{2}}$ and $G_{\theta_{3}}$.
Note also that while only four inchworm times $\theta_j$ are used, the Green's function is evaluated on a much finer interpolation grid.

\begin{figure}[bth]
    \includegraphics[width=\linewidth]{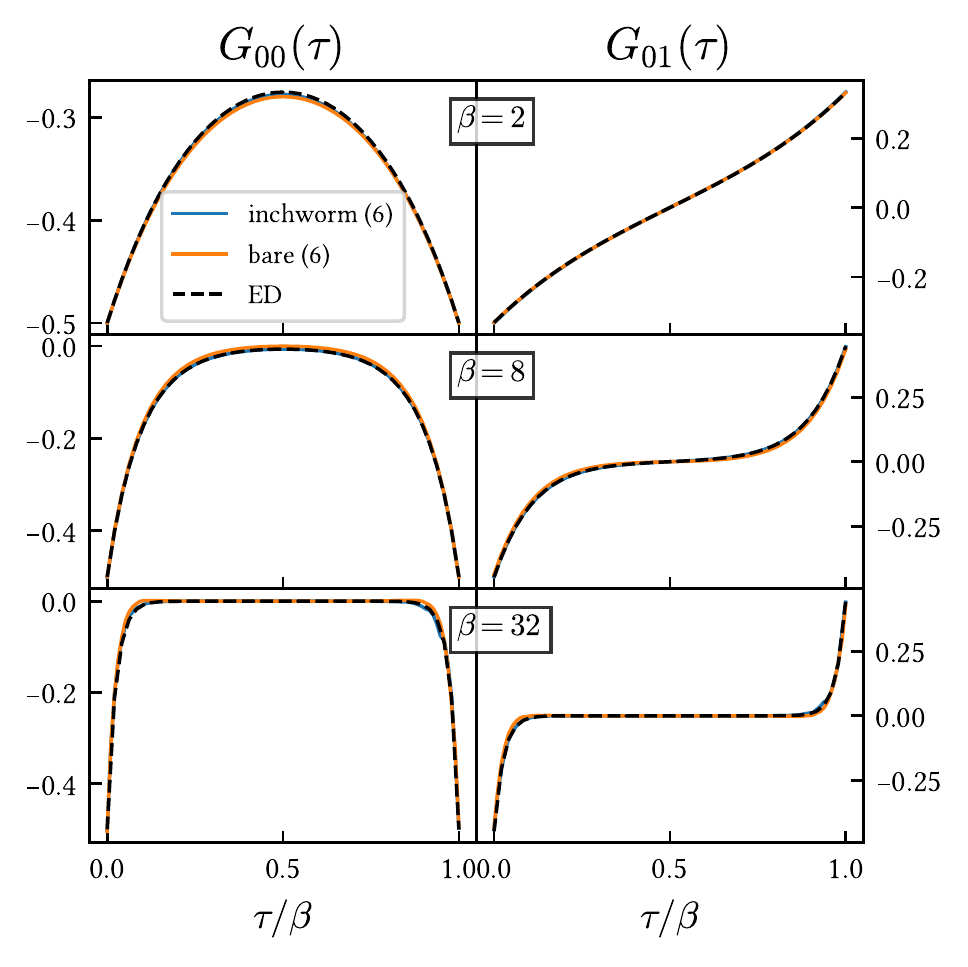}
    \caption{Inchworm results for a Hubbard dimer at half-filling using Hartree-shift at
        different temperatures, $U=2$, $k_{\max}=6$.
        Results are compared with ED and bare DiagMC results.
        Left (right) column shows the diagonal (off-diagonal) components.
    }
    \label{fig:dimer_hartree}
\end{figure}
Fig.~\ref{fig:dimer_hartree} shows the result of three methods for the Hubbard dimer (the two-site Hubbard model) at half-filling and $U=2$, at three inverse temperatures: $\beta=2$, $\beta=8$, and $\beta=32$.
We show the exact results obtained from ED; bare DiagMC results obtained by summing the first six orders in perturbation theory and truncating all remaining terms; and the inchworm Monte Carlo result, where each inchworm step is summed up to sixth order.
The two perturbation series are performed around the noninteracting system at the mean field level, which already includes the Hartree correction.

The results from ED (dashed black) and bare DiagMC (orange) differ slightly, indicating that the bare diagrammatic series is convergent---nevertheless, diagrams in the bare expansion beyond sixth order are not entirely negligible.
Inchworm results at the same order are in better agreement with ED, indicating somewhat faster convergence of the resummed series.

We employed a constant $\Delta \theta$, resulting in approximately linear increase of computer time with inverse temperature.
This is better than the typical case for CT-QMC methods, which scale cubically in the absence of a sign problem \cite{Gull11_RMP}.
An exponential increase of complexity, such as the need to go to higher diagram truncation orders when temperature is lowered, is not observed here.

\begin{figure}[bth]
    \includegraphics[width=\linewidth]{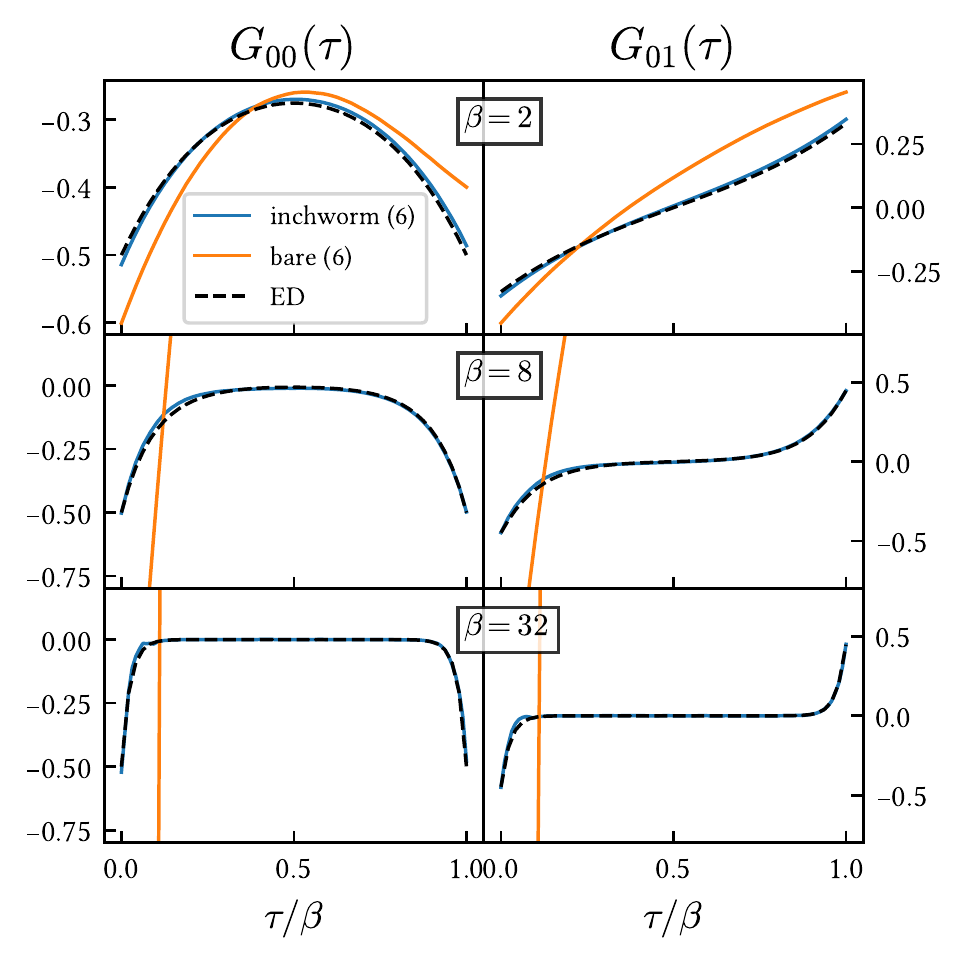}
    \caption{Inchworm results for a Hubbard dimer at half-filling without Hartree-shift at
        different temperatures, $U=2$, $k_{\max}=6$.
        Results are compared with ED and bare DiagMC results.
        Left (right) column shows the diagonal (off-diagonal) components.}
    \label{fig:dimer_no_hartree}
\end{figure}

Inchworm series converge more rapidly due to the renormalized propagators they employ.
One might therefore expect that the method should become more advantageous when the bare series diverges.
Since the convergence behavior of the diagrammatic series depends on the starting point, we remove the Hartree correction from the noninteracting starting point for the same system as in Fig.~\ref{fig:dimer_hartree}, and the results are shown Fig.~\ref{fig:dimer_no_hartree}.
For high temperatures ($\beta=2$, top panel) the bare perturbative series is not converged by order six.
For lower temperatures, signatures of a divergence in the bare series are visible.

In contrast, the inchworm series remains convergent for all parameter ranges studied here, and yields answers that are reasonably close to the exact result.
We attribute the remaining discrepancies between the converged inchworm solution and the exact solution to the order truncation, as well as the effect of stochastic noise from the Monte Carlo procedure.

\begin{figure}
    \includegraphics[width=\linewidth]{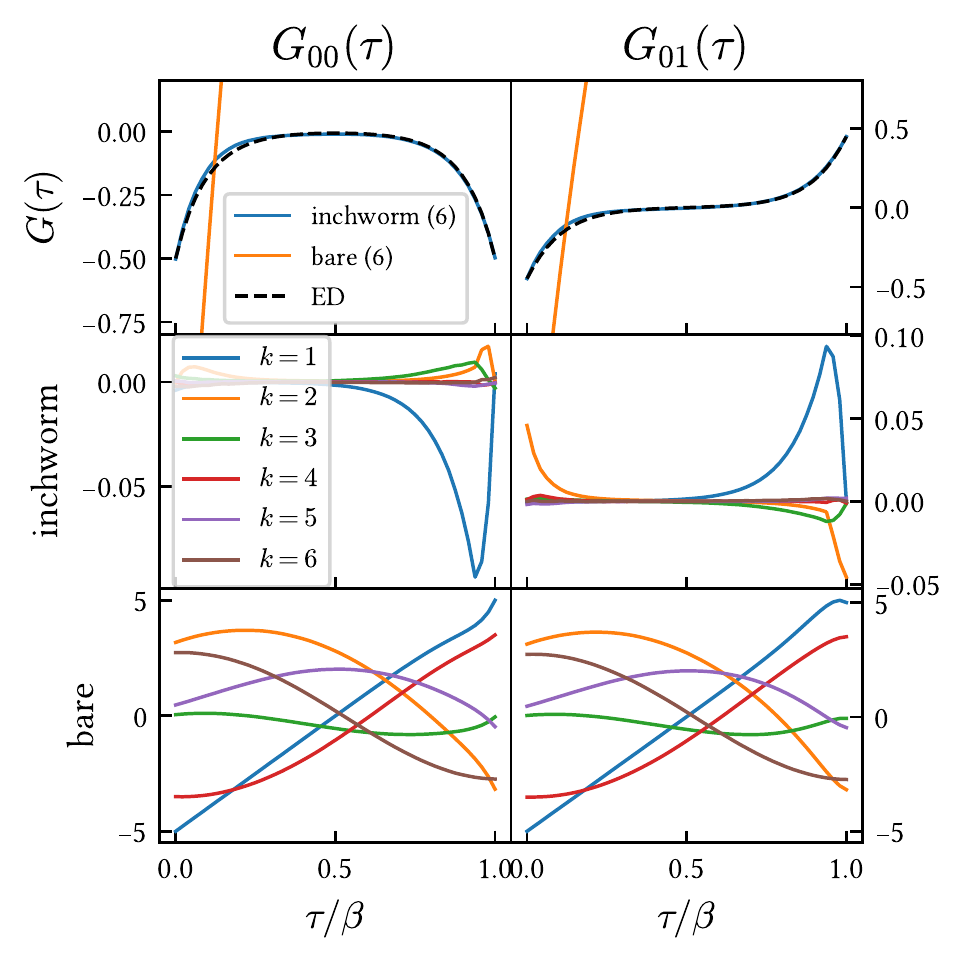}
    \caption{Convergence comparison between bare and inchworm Monte Carlo for the Hubbard dimer.
        Top panel shows the same results as in the middle panel of Fig.~\ref{fig:dimer_no_hartree}.
        Middle panel shows contributions from each expansion order in the final step of the
        inchworm calculation.
        Bottom panel shows corresponding order contributions in the bare DiagMC result.}
    \label{fig:dimer_compare}
\end{figure}

To further analyze the effect of order truncation, we disentangle the contributions from each diagram order to the final result in Fig.~\ref{fig:dimer_compare}.
The top two panels show the diagonal and off-diagonal Green's functions as discussed in Fig.~\ref{fig:dimer_no_hartree}.
The middle panels show the order-by-order contribution of the inchworm simulation to the final result.
The bottom panel shows the order-by-order contribution of the bare series.

Evidently, in the inchworm expansion the magnitude of contributions decreases rapidly with increasing order.
This is in sharp contrast to the bare result, where contributions grow with order.
Any resummation of these results would rely on cancellations between these contributions in order to obtain a converged result.

\begin{figure}[tb]
    \includegraphics[width=\linewidth]{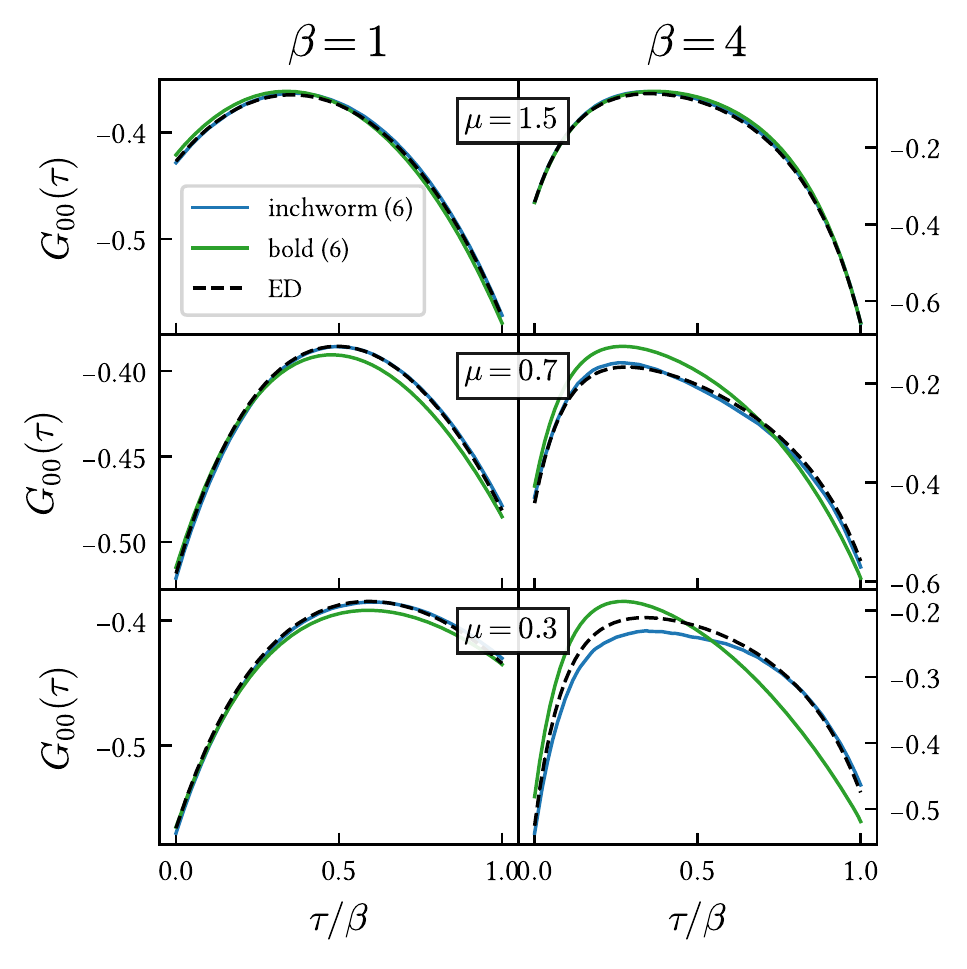}
    \caption{Inchworm results for a triangular Hubbard cluster at different temperatures and
        chemical potentials at $U=2$, $k_{\max}=6$, compared to bold DiagMC results and ED.
        Each row (column) corresponds to a different value of $\mu$ ($\beta$).
        Only the diagonal elements of the Green's function are shown.
    }
    \label{fig:trimer}
\end{figure}

Next, we compare the performance of the inchworm algorithm to the bold diagrammatic method~\cite{Prokofev2008b}. In that method, a self-energy is estimated via the summation of skeleton diagrams in terms of an approximate Green's function. The Dyson equation then provides an improved estimate of the Green's function, which is used to improve the guess for the self-energy, until both self-energy and green's function are self-consistent. The method is known to encounter difficulties, such as the convergence to unphysical fixed points, in areas where multiple self-consistent solutions exist~\cite{Kozik2015}.

As a test case we use the three-site periodic Hubbard chain with on-site interaction $U=2$. Fig.~\ref{fig:trimer} shows results for the on-site Green's function from ED, bold DiagMC, and inchworm, for two temperatures (left column: $\beta=1$; right column: $\beta=4$) and three values of the chemical potential. It is evident that bold-line Monte Carlo does not converge to the right result for all parameters shown. This behavior is caused by a truncation of the bold series at order $6$, and we expect that a higher diagram order would eventually lead to convergence.

In contrast, inchworm results for the same expansion order are well-converged for all cases except $\mu=0.3, \beta=4$.
While the primary discrepancy at these values comes from the truncation of the series, the effect of Monte Carlo noise is also clearly visible.
The fact that inchworm converges while the bold-line sampling does not shows that the two sampling procedures are very different, even though a precise connection between the inchworm expansion to the skeleton series can be made (see Appendix \ref{app:skeleton}).
In contrast to the bold algorithm, which `dresses' the propagator lines via a self-consistent iteration, inchworm dresses the propagator incrementally with well-defined auxiliary Green's functions at each iteration, and thus does not suffer from the misleading convergence problem
as reported in Ref.~\onlinecite{Kozik2015}.
However, because $G_\theta$ breaks time-translational symmetry,
such a symmetry breakage could persist in the final inchworm result in the presence of large order truncation errors.

\begin{figure}[tb]
    \includegraphics[width=\linewidth]{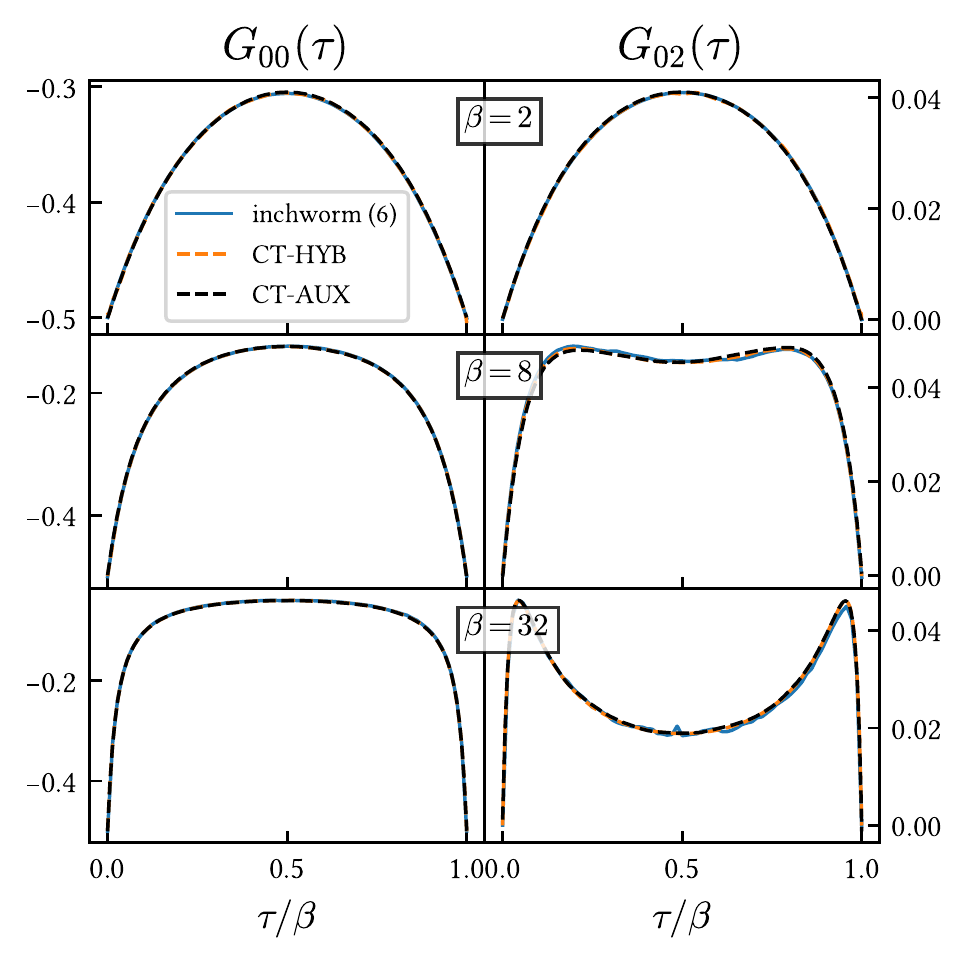}
    \caption{Inchworm results for a two-site Anderson impurity model with an off-diagonal hybridization at different temperatures, $U=2$, $k_{\max}=6$, $r=0.5$. Results are compared with CT-HYB and CT-AUX results. Left (right) column shows the diagonal (off-diagonal) components.
    }
    \label{fig:2AIM}
\end{figure}

Next, we test the inchworm method for a two-site quantum impurity problem with off-diagonal hybridizations.
Fig.~\ref{fig:2AIM} shows a comparison to CT-HYB and CT-AUX \cite{Gull08,Gull11_RMP,Gull11_submatrix} for a problem with hybridization function $\Delta_{i j}={\left[\delta_{i j}+r\left(1-\delta_{i j}\right)\right]} t^{2} D(\omega),$ $D(\omega)=1/\left( 2 \pi  t^2 \right) \sqrt{4 t^{2}-\omega^{2}}$.
We emphasize that the retardation effects of the bath in the impurity model are encapsulated in the bare impurity propagators, such that no explicit bath discretization is needed.
All algorithmic steps are therefore identical to the case of a lattice model, as in interaction expansion QMC \cite{Rubtsov05}.

Convergence in the diagonal and the off-diagonal Green's functions of the model is observed (within errors) between all methods within the six inchworm orders we employed here.
The inchworm result shows no systematic trend of deviation as temperature is lowered, although it can be observed that the statistical error becomes larger, due to the increasing number of inchworm steps required.
For the parameters explored here, both established algorithms (CT-HYB and CT-AUX) provide substantially more accurate results than the inchworm method for the same amount of computer time.

\begin{figure}[tb]
    \includegraphics[width=\linewidth]{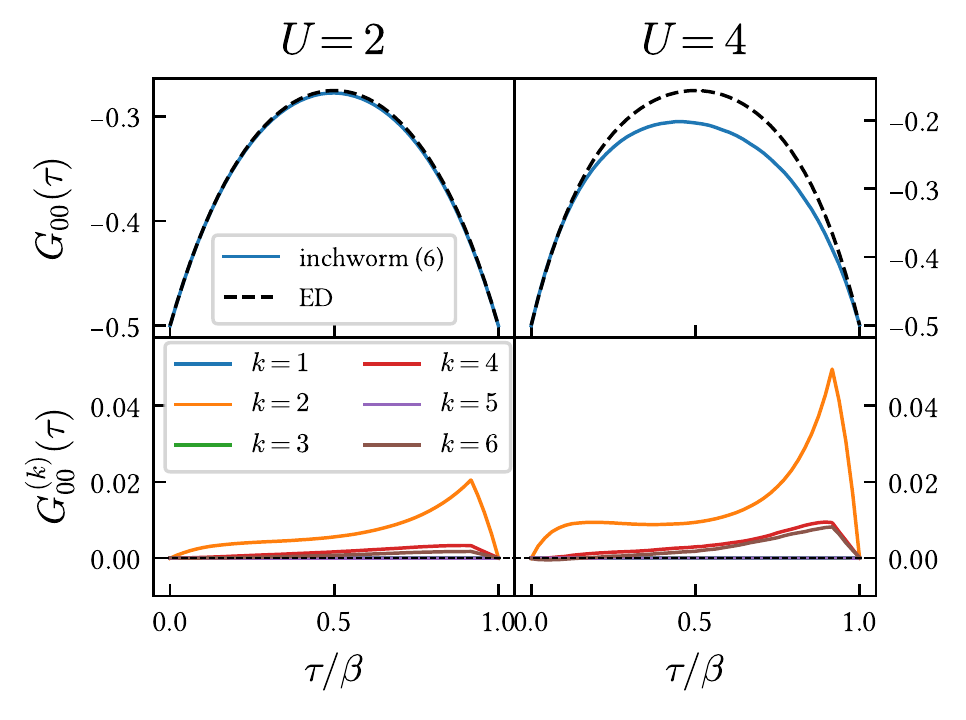}
    \caption{Inchworm results for a Hubbard dimer at half-filling with Hartree-shift with different interaction $U$ at $\beta=2$, $k_{\max}=6$.
        Top row shows results for diagonal components of the Green's function compared with ED results.
        Bottom row shows the order-by-order contribution to the final Green's function.
    }
    \label{fig:fail}
\end{figure}

Finally, for stronger interactions, the inchworm algorithm requires more expansion orders at each step to obtain converged results.
As Fig.~\ref{fig:fail} shows for the Hubbard dimer, increasing the interaction strength typically also increases the contributions from higher orders.
The truncation errors associated with neglecting high orders cannot be controlled by simply decreasing $\Delta\theta$ (see Appendix~\ref{app:skeleton} for an analysis at $\Delta\theta\to 0$).
This is the major limitation of the inchworm interaction method, and is due to the perturbative nature of the formalism.
Nevertheless, with faster series convergence (compared with bare DiagMC) and no instability due to self-consistency (compared to bold DiagMC), the inchworm interaction algorithm provides an alternative path in the development of DiagMC methods.

\section{Conclusion}\label{sec:conclusions}
In conclusion, we have shown that the idea of inchworm expansions, originally applied to the hybridization expansion for quantum impurity models, is also relevant to interaction expansions.
The method is shown to converge in regimes where bare Monte Carlo diverges, and is shown to give the correct answer at low order in regimes where bold-line Monte Carlo is observed to converge to an incorrect result at the same expansion order; the connection between the (iterative) summation of the inchworm series and the (self-consistent) summation of the skeleton series is discussed in Appendix~\ref{app:skeleton}.
An explicit bath discretization, such as needed in ED or wavefunction-based quantum chemistry approaches, is not needed.

We considered applications of the method to very strongly correlated impurity systems, such as those typically employed within dynamical mean field theory and self-energy embedding theory.
We found that for typical applications of quantum impurity solvers within these domains, the interaction inchworm method is not yet competitive with established CT-QMC techniques like CT-HYB and CT-AUX.
Changes in this assessment may develop if improvements to the algorithm are implemented: for example, fast diagram summation schemes \cite{Rossi2017b,Boag2018} could enable the method to reach much higher orders.
Looking forward, however, we believe the main advantages of the interaction inchworm method will come into play when we begin to take advantage of its flexibility to inch in \emph{space} rather than \emph{time}.
This will allow us to use it in conjunction with other impurity solvers, potentially resulting in a powerful new set of tools.

\begin{acknowledgments}
    JL was supported by the Simons foundation via the Simons collaboration on the many-electron problem. YY was supported by NSF under NSF-DMR 2001465. EG was supported by Department of Energy via grant DESC0022088.
    G.C. acknowledges support by the Israel Science Foundation (Grants No. 2902/21 and 218/19) and by the PAZY foundation (Grant No. 308/19).
\end{acknowledgments}

\appendix

\section{Explicit definition of the auxiliary Green's function}\label{app:Gaux-explicit}
Introducing
\begin{equation}
    \hat{S}_I(\tau,\tau') = \hat{U}_I(\tau) \hat{U}_I^{-1}(\tau'),
\end{equation}
we can write the physical Green's function $G$ in interaction picture as
\begin{widetext}
    \begin{equation}
        G(\tau,\tau')
        \quad=
        \begin{cases}
            -\frac{1}{Z}\Tr[e^{-\beta\Hhat_{0}}\hat{S}_I(\beta,\tau)\hat{c}_{I}(\tau)\hat{S}_I(\tau,\tau')\hat{c}_{I}^{\dagger}(\tau')\hat{S}_I(\tau',0)]: & \tau > \tau', \\
            \frac{1}{Z}\Tr[e^{-\beta\Hhat_{0}}\hat{S}_I(\beta,\tau')\hat{c}_{I}^{\dagger}(\tau')\hat{S}_I(\tau',\tau)\hat{c}_{I}(\tau)\hat{S}_I(\tau,0)]:  & \tau < \tau'.
        \end{cases}
    \end{equation}
The auxiliary Green's function can be formulated in a similar manner: \begin{equation}
        G_{\theta}(\tau,\tau')=\begin{cases}
            -\frac{1}{Z_{\theta}}\Tr[e^{-\beta\Hhat_{0}}\hat{c}_{I}(\tau)\hat{c}_{I}^{\dagger}(\tau')\hat{S}_I(\theta,0)],                                           & \tau>\tau'>\theta, \\
            \frac{1}{Z_{\theta}}\Tr[e^{-\beta\Hhat_{0}}\hat{c}_{I}^{\dagger}(\tau')\hat{c}_{I}(\tau)\hat{S}_I(\theta,0)],                                            & \tau'>\tau>\theta, \\
            -\frac{1}{Z_{\theta}}\Tr[e^{-\beta\Hhat_{0}}\hat{c}_{I}(\tau)\hat{S}_I(\theta,\tau')\hat{c}_{I}^{\dagger}(\tau')\hat{U}_{I}(\tau')],                     & \tau>\theta>\tau', \\
            \frac{1}{Z_{\theta}}\Tr[e^{-\beta\Hhat_{0}}\hat{c}_{I}^{\dagger}(\tau')\hat{S}_I(\theta,\tau)\hat{c}_{I}(\tau)\hat{U}_{I}(\tau)],                        & \tau'>\theta>\tau, \\
            -\frac{1}{Z_{\theta}}\Tr[e^{-\beta\Hhat_{0}}\hat{S}_I(\theta,\tau)\hat{c}_{I}(\tau)\hat{S}_I(\tau,\tau')\hat{c}_{I}^{\dagger}(\tau')\hat{U}_{I}(\tau')], & \theta>\tau>\tau', \\
            \frac{1}{Z_{\theta}}\Tr[e^{-\beta\Hhat_{0}}\hat{S}_I(\theta,\tau')\hat{c}_{I}^{\dagger}(\tau')\hat{S}_I(\tau',\tau)\hat{c}_{I}(\tau)\hat{U}_{I}(\tau)],  & \theta>\tau'>\tau.
        \end{cases}
    \end{equation}
    One can verify that this is equivalent to Eq.~(\ref{eq:Gaux-expansion}) by
    plugging in the Dyson series for $\hat{S}_I$~\cite{Mahan2000}:
    \begin{equation}
        \hat{S}_I(\tau,\tau') = \sum_{k=0}^{\infty}\frac{(-1)^{k}}{k!}\int_{\tau'}^{\tau}\drm\tau_{1}\int_{\tau'}^{\tau}\drm\tau_{2}
        \cdots\int_{\tau'}^{\tau}\drm\tau_{k}
        \Tcal_{\tau}\{\hat{V}_{I}(\tau_{1})\hat{V}_{I}(\tau_{2})\cdots\hat{V}_{I}(\tau_{k})\}.
    \end{equation}
    Expanding all interaction picture operators explicitly, we have
    \begin{equation}
        G_{\theta}(\tau,\tau')=\begin{cases}
            -\frac{1}{Z_{\theta}}\Tr[e^{-(\beta-\tau)\Hhat_{0}}\hat{c}e^{-(\tau-\tau')\Hhat_{0}}\hat{c}^{\dagger}e^{-(\tau'-\theta)\Hhat_{0}}e^{-\theta\Hhat}], & \tau>\tau'>\theta, \\
            \frac{1}{Z_{\theta}}\Tr[e^{-(\beta-\tau')\Hhat_{0}}\hat{c}^{\dagger}e^{-(\tau'-\tau)\Hhat_{0}}\hat{c}e^{-(\tau-\theta)\Hhat_{0}}e^{-\theta\Hhat}],  & \tau'>\tau>\theta, \\
            -\frac{1}{Z_{\theta}}\Tr[e^{-(\beta-\tau)\Hhat_{0}}\hat{c}e^{-(\tau-\theta)\Hhat_{0}}e^{-(\theta-\tau')\Hhat}\hat{c}^{\dagger}e^{-\tau'\Hhat}],     & \tau>\theta>\tau', \\
            \frac{1}{Z_{\theta}}\Tr[e^{-(\beta-\tau')\Hhat_{0}}\hat{c}^{\dagger}e^{-(\tau'-\theta)\Hhat_{0}}e^{-(\theta-\tau)\Hhat}\hat{c}e^{-\tau\Hhat}],      & \tau'>\theta>\tau, \\
            -\frac{1}{Z_{\theta}}\Tr[e^{-(\beta-\theta)\Hhat_{0}}e^{-(\theta-\tau)\Hhat}\hat{c}e^{-(\tau-\tau')\Hhat}\hat{c}^{\dagger}e^{-\tau'\Hhat}],         & \theta>\tau>\tau', \\
            \frac{1}{Z_{\theta}}\Tr[e^{-(\beta-\theta)\Hhat_{0}}e^{-(\theta-\tau')\Hhat}\hat{c}^{\dagger}e^{-(\tau'-\tau)\Hhat}\hat{c}e^{-\tau\Hhat}],          & \theta>\tau'>\tau,
        \end{cases}
    \end{equation}
    which can be used to compute $G_\theta$ numerically by exact diagonalization for small models.
\end{widetext}

\section{Connection between the inchworm expansion and the skeleton series}\label{app:skeleton}

The diagram rules for the inchworm expansion are reminiscent of the skeleton
diagram rules~\cite{Luttinger1960} due to the exclusion of two-particle reducible Type 1 components.
The connection between the inchworm expansion (\ref{eq:formal-inch-expansion}) and
the skeleton series can be revealed in the limit where $\theta'=\beta$,
$\theta=\beta-\Delta\theta$, and $\Delta\theta\to 0$.
For convenience,
rewrite
Eq.~(\ref{eq:Zaux-expansion}) as a coherent state path integral~\cite{Negele1988}
\begin{equation}
\begin{split}
    Z_{\theta}  =\int\Dcal[\bar{c},c]e^{-S_{0}}\exp\left(-\int_{0}^{\theta}\drm\tau V(\tau)\right),
\end{split}\label{eq:Zaux-action}
\end{equation}
where $\bar{c}(\tau)$ and $c(\tau)$ are Grassmann fields, $S_0$ the non-interacting action,
and $V(\tau)$ the Grassmann function obtained by replacing operators
$\hat{c}^\dagger$ and $\hat{c}$ in $\Vhat$ with the Grassmann fields.
The generating function $\mathcal{W}_\theta$ of the auxiliary Green's function is the logarithm of
$Z_\theta$ with a bilinear source term $J$~\cite{Negele1988}:
\begin{equation}
    \begin{aligned}
        Z_\theta[J] &=\int\Dcal[\bar{c},c]\exp\bigg(-S_0-\int_{0}^{\theta}\drm\tau V(\tau)\\
        &+\int_0^\beta\drm\tau'\drm\tau \bar{c}(\tau')J(\tau',\tau)c(\tau) \bigg),\\
        \mathcal{W}_\theta[J] &:= \log Z_\theta[J],\qquad
        \frac{\delta \mathcal{W}_\theta}{\delta J(\tau',\tau)}\bigg|_{J=0} = G_\theta(\tau,\tau').
    \end{aligned}
\end{equation}
When $\Delta\theta\to 0$, we have
\begin{equation}
    G_\theta - G_{\theta-\Delta\theta} \approx \frac{\partial G_\theta}{\partial\theta} \Delta\theta
    = \frac{\delta}{\delta J} \frac{\partial \mathcal{W}_\theta}{\partial\theta}\bigg|_{J=0}\Delta\theta.
\end{equation}
If the series expansion of $\partial_\theta \mathcal{W}_\theta$ uniformly
converges near $J = 0$, its derivative is also expected to converge.
The convergence of this infinitesimal inchworm expansion for $G_\theta$ is thus
directly related to the convergence properties of
$\partial_\theta\mathcal{W}_\theta |_{J=0} = \partial_\theta\log Z_\theta$.

From (\ref{eq:Zaux-action}), we have
\begin{equation}
    \begin{aligned}
        \frac{\partial}{\partial\theta}\log Z_\theta &= \frac{1}{Z_\theta}\frac{\partial Z_\theta}{\partial\theta} =
        \frac{1}{Z_\theta} \int\Dcal[\bar{c},c](-V(\theta))e^{-S_{0}}\\
        &\times\exp\bigg(-\int_{0}^{\theta}\drm\tau V(\tau)\bigg).
    \end{aligned}
\end{equation}
When taking $\theta=\beta$, $Z_\theta$ becomes $Z$, and the integral of
$V(\tau)$ recovers the interacting action $S_V$ and we have
\begin{equation}
\begin{split}
    &\frac{\partial}{\partial\theta}\log Z_\theta\bigg|_{\theta=\beta} =
    \frac{1}{Z} \int\Dcal[\bar{c},c](-V(\beta))e^{-S} = -\langle\Vhat\rangle,
\end{split}\label{eq:logZ}
\end{equation}
where $S=S_0+S_V$ is the full action of the system.
For a standard four-fermion interaction $\Vhat$, the expectation value can be
formulated in terms of the Green's function $G$ and the self-energy $\Sigma$
\begin{equation}
    \langle\Vhat\rangle = \frac{1}{2\beta}\sum_{n}\Tr [\Sigma(\iwn)G(\iwn)].
    \label{eq:TrSigmaG}
\end{equation}
$\Sigma$ can be obtained as a functional derivative of the Luttinger-Ward functional
$\Phi[G]$~\cite{Luttinger1960}: \begin{equation}
    \frac{\delta\Phi}{\delta G} = \Sigma[G],
\end{equation}
and the skeleton series can be formally written as~\cite{Luttinger1960,Lin2018}
\begin{equation}
    \begin{aligned}
   \Sigma[G]  &= \sum_{k=1}^\infty \Sigma^{(k)}[G],\quad \Phi[G] = \sum_{k=1}^\infty \Phi^{(k)}[G],\\
   \Phi^{(k)}  &= \frac{1}{2k} \Tr[\Sigma^{(k)} G] = \frac{1}{2k} \sum_{n}\Tr[\Sigma^{(k)}(\iwn) G(\iwn)],
    \end{aligned}\label{eq:skeleton-series}
\end{equation}
where $\Sigma^{(k)}$ is the sum of all $k$-th order skeleton diagrams.
Combining Eqs.~(\ref{eq:logZ}), (\ref{eq:TrSigmaG}), and
(\ref{eq:skeleton-series}), we have
\begin{equation}
\begin{aligned}
    \frac{\partial}{\partial\theta}\log Z_\theta\bigg|_{\theta=\beta} &=
    -\sum_{k=1}^\infty \frac{1}{2\beta}\sum_{n}\Tr [\Sigma^{(k)}(\iwn)G(\iwn)]\\
    &= -\frac{1}{\beta}\sum_{k=1}^\infty k\Phi^{(k)}.
\end{aligned}\label{eq:logZ-series}
\end{equation}
This directly relates the inchworm expansion at $\theta=\beta$ to the skeleton expansion of the Luttinger-Ward functional.
If the skeleton series (\ref{eq:skeleton-series}) is absolutely convergent, so
should Eq.~(\ref{eq:logZ-series}), which implies a convergent inchworm expansion
at $G_{\theta=\beta}$.

\bibliographystyle{apsrev4-2}

\begin{thebibliography}{56}\makeatletter
\providecommand \@ifxundefined [1]{\@ifx{#1\undefined}
}\providecommand \@ifnum [1]{\ifnum #1\expandafter \@firstoftwo
 \else \expandafter \@secondoftwo
 \fi
}\providecommand \@ifx [1]{\ifx #1\expandafter \@firstoftwo
 \else \expandafter \@secondoftwo
 \fi
}\providecommand \natexlab [1]{#1}\providecommand \enquote  [1]{``#1''}\providecommand \bibnamefont  [1]{#1}\providecommand \bibfnamefont [1]{#1}\providecommand \citenamefont [1]{#1}\providecommand \href@noop [0]{\@secondoftwo}\providecommand \href [0]{\begingroup \@sanitize@url \@href}\providecommand \@href[1]{\@@startlink{#1}\@@href}\providecommand \@@href[1]{\endgroup#1\@@endlink}\providecommand \@sanitize@url [0]{\catcode `\\12\catcode `\$12\catcode
  `\&12\catcode `\#12\catcode `\^12\catcode `\_12\catcode `\%12\relax}\providecommand \@@startlink[1]{}\providecommand \@@endlink[0]{}\providecommand \url  [0]{\begingroup\@sanitize@url \@url }\providecommand \@url [1]{\endgroup\@href {#1}{\urlprefix }}\providecommand \urlprefix  [0]{URL }\providecommand \Eprint [0]{\href }\providecommand \doibase [0]{https://doi.org/}\providecommand \selectlanguage [0]{\@gobble}\providecommand \bibinfo  [0]{\@secondoftwo}\providecommand \bibfield  [0]{\@secondoftwo}\providecommand \translation [1]{[#1]}\providecommand \BibitemOpen [0]{}\providecommand \bibitemStop [0]{}\providecommand \bibitemNoStop [0]{.\EOS\space}\providecommand \EOS [0]{\spacefactor3000\relax}\providecommand \BibitemShut  [1]{\csname bibitem#1\endcsname}\let\auto@bib@innerbib\@empty
\bibitem [{\citenamefont {Georges}\ \emph {et~al.}(1996)\citenamefont
  {Georges}, \citenamefont {Kotliar}, \citenamefont {Krauth},\ and\
  \citenamefont {Rozenberg}}]{Georges96}\BibitemOpen
  \bibfield  {author} {\bibinfo {author} {\bibfnamefont {A.}~\bibnamefont
  {Georges}}, \bibinfo {author} {\bibfnamefont {G.}~\bibnamefont {Kotliar}},
  \bibinfo {author} {\bibfnamefont {W.}~\bibnamefont {Krauth}},\ and\ \bibinfo
  {author} {\bibfnamefont {M.~J.}\ \bibnamefont {Rozenberg}},\ }\href
  {https://doi.org/10.1103/RevModPhys.68.13} {\bibfield  {journal} {\bibinfo
  {journal} {Rev. Mod. Phys.}\ }\textbf {\bibinfo {volume} {68}},\ \bibinfo
  {pages} {13} (\bibinfo {year} {1996})}\BibitemShut {NoStop}\bibitem [{\citenamefont {Kotliar}\ \emph {et~al.}(2006)\citenamefont
  {Kotliar}, \citenamefont {Savrasov}, \citenamefont {Haule} \emph
  {et~al.}}]{Kotliar06}\BibitemOpen
  \bibfield  {author} {\bibinfo {author} {\bibfnamefont {G.}~\bibnamefont
  {Kotliar}}, \bibinfo {author} {\bibfnamefont {S.~Y.}\ \bibnamefont
  {Savrasov}}, \bibinfo {author} {\bibfnamefont {K.}~\bibnamefont {Haule}},
  \emph {et~al.},\ }\href {https://doi.org/10.1103/RevModPhys.78.865}
  {\bibfield  {journal} {\bibinfo  {journal} {Rev. Mod. Phys.}\ }\textbf
  {\bibinfo {volume} {78}},\ \bibinfo {eid} {865} (\bibinfo {year}
  {2006})}\BibitemShut {NoStop}\bibitem [{\citenamefont {Zgid}\ and\ \citenamefont {Gull}(2017)}]{Zgid17}\BibitemOpen
  \bibfield  {author} {\bibinfo {author} {\bibfnamefont {D.}~\bibnamefont
  {Zgid}}\ and\ \bibinfo {author} {\bibfnamefont {E.}~\bibnamefont {Gull}},\
  }\href {https://doi.org/10.1088/1367-2630/aa5d34} {\bibfield  {journal}
  {\bibinfo  {journal} {New Journal of Physics}\ }\textbf {\bibinfo {volume}
  {19}},\ \bibinfo {pages} {023047} (\bibinfo {year} {2017})}\BibitemShut
  {NoStop}\bibitem [{\citenamefont {Brako}\ and\ \citenamefont {Newns}(1981)}]{Brako81}\BibitemOpen
  \bibfield  {author} {\bibinfo {author} {\bibfnamefont {R.}~\bibnamefont
  {Brako}}\ and\ \bibinfo {author} {\bibfnamefont {D.~M.}\ \bibnamefont
  {Newns}},\ }\href {https://doi.org/10.1088/0022-3719/14/21/023} {\bibfield
  {journal} {\bibinfo  {journal} {Journal of Physics C: Solid State Physics}\
  }\textbf {\bibinfo {volume} {14}},\ \bibinfo {pages} {3065} (\bibinfo {year}
  {1981})}\BibitemShut {NoStop}\bibitem [{\citenamefont {Langreth}\ and\ \citenamefont
  {Nordlander}(1991)}]{Langreth91}\BibitemOpen
  \bibfield  {author} {\bibinfo {author} {\bibfnamefont {D.~C.}\ \bibnamefont
  {Langreth}}\ and\ \bibinfo {author} {\bibfnamefont {P.}~\bibnamefont
  {Nordlander}},\ }\href {https://doi.org/10.1103/PhysRevB.43.2541} {\bibfield
  {journal} {\bibinfo  {journal} {Phys. Rev. B}\ }\textbf {\bibinfo {volume}
  {43}},\ \bibinfo {pages} {2541} (\bibinfo {year} {1991})}\BibitemShut
  {NoStop}\bibitem [{\citenamefont {Anderson}(1961)}]{Anderson61}\BibitemOpen
  \bibfield  {author} {\bibinfo {author} {\bibfnamefont {P.~W.}\ \bibnamefont
  {Anderson}},\ }\href {https://doi.org/10.1103/PhysRev.124.41} {\bibfield
  {journal} {\bibinfo  {journal} {Phys. Rev.}\ }\textbf {\bibinfo {volume}
  {124}},\ \bibinfo {pages} {41} (\bibinfo {year} {1961})}\BibitemShut
  {NoStop}\bibitem [{\citenamefont {Datta}(1997)}]{datta_electronic_1997}\BibitemOpen
  \bibfield  {author} {\bibinfo {author} {\bibfnamefont {S.}~\bibnamefont
  {Datta}},\ }\href@noop {} {\emph {\bibinfo {title} {Electronic {{Transport}}
  in {{Mesoscopic Systems}}}}}\ (\bibinfo  {publisher} {{Cambridge
  UniversityPress}},\ \bibinfo {year} {1997})\BibitemShut {NoStop}\bibitem [{\citenamefont {Hanson}\ \emph {et~al.}(2007)\citenamefont {Hanson},
  \citenamefont {Kouwenhoven}, \citenamefont {Petta}, \citenamefont {Tarucha},\
  and\ \citenamefont {Vandersypen}}]{Hanson07}\BibitemOpen
  \bibfield  {author} {\bibinfo {author} {\bibfnamefont {R.}~\bibnamefont
  {Hanson}}, \bibinfo {author} {\bibfnamefont {L.~P.}\ \bibnamefont
  {Kouwenhoven}}, \bibinfo {author} {\bibfnamefont {J.~R.}\ \bibnamefont
  {Petta}}, \bibinfo {author} {\bibfnamefont {S.}~\bibnamefont {Tarucha}},\
  and\ \bibinfo {author} {\bibfnamefont {L.~M.~K.}\ \bibnamefont
  {Vandersypen}},\ }\href {https://doi.org/10.1103/RevModPhys.79.1217}
  {\bibfield  {journal} {\bibinfo  {journal} {Reviews of Modern Physics}\
  }\textbf {\bibinfo {volume} {79}},\ \bibinfo {eid} {1217} (\bibinfo {year}
  {2007})}\BibitemShut {NoStop}\bibitem [{\citenamefont {Aviram}\ and\ \citenamefont
  {Ratner}(1974)}]{aviram_molecular_1974}\BibitemOpen
  \bibfield  {author} {\bibinfo {author} {\bibfnamefont {A.}~\bibnamefont
  {Aviram}}\ and\ \bibinfo {author} {\bibfnamefont {M.~A.}\ \bibnamefont
  {Ratner}},\ }\href@noop {} {\bibfield  {journal} {\bibinfo  {journal} {Chem.
  Phys. Lett.}\ }\textbf {\bibinfo {volume} {29}},\ \bibinfo {pages} {277}
  (\bibinfo {year} {1974})}\BibitemShut {NoStop}\bibitem [{\citenamefont {Nitzan}\ and\ \citenamefont
  {Ratner}(2003)}]{nitzan_electron_2003}\BibitemOpen
  \bibfield  {author} {\bibinfo {author} {\bibfnamefont {A.}~\bibnamefont
  {Nitzan}}\ and\ \bibinfo {author} {\bibfnamefont {M.~A.}\ \bibnamefont
  {Ratner}},\ }\href {https://doi.org/10.1126/science.1081572} {\bibfield
  {journal} {\bibinfo  {journal} {Science}\ }\textbf {\bibinfo {volume}
  {300}},\ \bibinfo {pages} {1384} (\bibinfo {year} {2003})}\BibitemShut
  {NoStop}\bibitem [{\citenamefont {Cohen}\ and\ \citenamefont
  {Galperin}(2020)}]{cohen_greens_2020}\BibitemOpen
  \bibfield  {author} {\bibinfo {author} {\bibfnamefont {G.}~\bibnamefont
  {Cohen}}\ and\ \bibinfo {author} {\bibfnamefont {M.}~\bibnamefont
  {Galperin}},\ }\href {https://doi.org/10.1063/1.5145210} {\bibfield
  {journal} {\bibinfo  {journal} {The Journal of Chemical Physics}\ }\textbf
  {\bibinfo {volume} {152}},\ \bibinfo {pages} {090901} (\bibinfo {year}
  {2020})}\BibitemShut {NoStop}\bibitem [{\citenamefont {Prokof'ev}\ \emph {et~al.}(1998)\citenamefont
  {Prokof'ev}, \citenamefont {Svistunov},\ and\ \citenamefont
  {Tupitsyn}}]{Prokofev98A}\BibitemOpen
  \bibfield  {author} {\bibinfo {author} {\bibfnamefont {N.~V.}\ \bibnamefont
  {Prokof'ev}}, \bibinfo {author} {\bibfnamefont {B.~V.}\ \bibnamefont
  {Svistunov}},\ and\ \bibinfo {author} {\bibfnamefont {I.~S.}\ \bibnamefont
  {Tupitsyn}},\ }\href {https://doi.org/10.1134/1.558661} {\bibfield  {journal}
  {\bibinfo  {journal} {Journal of Experimental and Theoretical Physics}\
  }\textbf {\bibinfo {volume} {87}},\ \bibinfo {pages} {310} (\bibinfo {year}
  {1998})}\BibitemShut {NoStop}\bibitem [{\citenamefont {Prokof'ev}\ and\ \citenamefont
  {Svistunov}(1998)}]{Prokofev98B}\BibitemOpen
  \bibfield  {author} {\bibinfo {author} {\bibfnamefont {N.~V.}\ \bibnamefont
  {Prokof'ev}}\ and\ \bibinfo {author} {\bibfnamefont {B.~V.}\ \bibnamefont
  {Svistunov}},\ }\href {https://doi.org/10.1103/PhysRevLett.81.2514}
  {\bibfield  {journal} {\bibinfo  {journal} {Physical Review Letters}\
  }\textbf {\bibinfo {volume} {81}},\ \bibinfo {pages} {2514} (\bibinfo {year}
  {1998})}\BibitemShut {NoStop}\bibitem [{\citenamefont {Caffarel}\ and\ \citenamefont
  {Krauth}(1994)}]{Caffarel94}\BibitemOpen
  \bibfield  {author} {\bibinfo {author} {\bibfnamefont {M.}~\bibnamefont
  {Caffarel}}\ and\ \bibinfo {author} {\bibfnamefont {W.}~\bibnamefont
  {Krauth}},\ }\href {https://doi.org/10.1103/PhysRevLett.72.1545} {\bibfield
  {journal} {\bibinfo  {journal} {Phys. Rev. Lett.}\ }\textbf {\bibinfo
  {volume} {72}},\ \bibinfo {pages} {1545} (\bibinfo {year}
  {1994})}\BibitemShut {NoStop}\bibitem [{\citenamefont {Iskakov}\ and\ \citenamefont
  {Danilov}(2018)}]{Iskakov18}\BibitemOpen
  \bibfield  {author} {\bibinfo {author} {\bibfnamefont {S.}~\bibnamefont
  {Iskakov}}\ and\ \bibinfo {author} {\bibfnamefont {M.}~\bibnamefont
  {Danilov}},\ }\href
  {https://doi.org/https://doi.org/10.1016/j.cpc.2017.12.016} {\bibfield
  {journal} {\bibinfo  {journal} {Computer Physics Communications}\ }\textbf
  {\bibinfo {volume} {225}},\ \bibinfo {pages} {128} (\bibinfo {year}
  {2018})}\BibitemShut {NoStop}\bibitem [{\citenamefont {Bulla}\ \emph {et~al.}(2008)\citenamefont {Bulla},
  \citenamefont {Costi},\ and\ \citenamefont {Pruschke}}]{Bulla08}\BibitemOpen
  \bibfield  {author} {\bibinfo {author} {\bibfnamefont {R.}~\bibnamefont
  {Bulla}}, \bibinfo {author} {\bibfnamefont {T.~A.}\ \bibnamefont {Costi}},\
  and\ \bibinfo {author} {\bibfnamefont {T.}~\bibnamefont {Pruschke}},\ }\href
  {https://doi.org/10.1103/RevModPhys.80.395} {\bibfield  {journal} {\bibinfo
  {journal} {Rev. Mod. Phys.}\ }\textbf {\bibinfo {volume} {80}},\ \bibinfo
  {pages} {395} (\bibinfo {year} {2008})}\BibitemShut {NoStop}\bibitem [{\citenamefont {Bulla}(1999)}]{Bulla99}\BibitemOpen
  \bibfield  {author} {\bibinfo {author} {\bibfnamefont {R.}~\bibnamefont
  {Bulla}},\ }\href {https://doi.org/10.1103/PhysRevLett.83.136} {\bibfield
  {journal} {\bibinfo  {journal} {Phys. Rev. Lett.}\ }\textbf {\bibinfo
  {volume} {83}},\ \bibinfo {pages} {136} (\bibinfo {year} {1999})}\BibitemShut
  {NoStop}\bibitem [{\citenamefont {Wolf}\ \emph {et~al.}(2014)\citenamefont {Wolf},
  \citenamefont {McCulloch}, \citenamefont {Parcollet},\ and\ \citenamefont
  {Schollw\"ock}}]{Wolf14}\BibitemOpen
  \bibfield  {author} {\bibinfo {author} {\bibfnamefont {F.~A.}\ \bibnamefont
  {Wolf}}, \bibinfo {author} {\bibfnamefont {I.~P.}\ \bibnamefont {McCulloch}},
  \bibinfo {author} {\bibfnamefont {O.}~\bibnamefont {Parcollet}},\ and\
  \bibinfo {author} {\bibfnamefont {U.}~\bibnamefont {Schollw\"ock}},\ }\href
  {https://doi.org/10.1103/PhysRevB.90.115124} {\bibfield  {journal} {\bibinfo
  {journal} {Phys. Rev. B}\ }\textbf {\bibinfo {volume} {90}},\ \bibinfo
  {pages} {115124} (\bibinfo {year} {2014})}\BibitemShut {NoStop}\bibitem [{\citenamefont {Bauernfeind}\ \emph {et~al.}(2017)\citenamefont
  {Bauernfeind}, \citenamefont {Zingl}, \citenamefont {Triebl}, \citenamefont
  {Aichhorn},\ and\ \citenamefont {Evertz}}]{Bauernfeind17}\BibitemOpen
  \bibfield  {author} {\bibinfo {author} {\bibfnamefont {D.}~\bibnamefont
  {Bauernfeind}}, \bibinfo {author} {\bibfnamefont {M.}~\bibnamefont {Zingl}},
  \bibinfo {author} {\bibfnamefont {R.}~\bibnamefont {Triebl}}, \bibinfo
  {author} {\bibfnamefont {M.}~\bibnamefont {Aichhorn}},\ and\ \bibinfo
  {author} {\bibfnamefont {H.~G.}\ \bibnamefont {Evertz}},\ }\href
  {https://doi.org/10.1103/PhysRevX.7.031013} {\bibfield  {journal} {\bibinfo
  {journal} {Phys. Rev. X}\ }\textbf {\bibinfo {volume} {7}},\ \bibinfo {pages}
  {031013} (\bibinfo {year} {2017})}\BibitemShut {NoStop}\bibitem [{\citenamefont {White}\ and\ \citenamefont
  {Feiguin}(2004)}]{Feiguin04}\BibitemOpen
  \bibfield  {author} {\bibinfo {author} {\bibfnamefont {S.~R.}\ \bibnamefont
  {White}}\ and\ \bibinfo {author} {\bibfnamefont {A.~E.}\ \bibnamefont
  {Feiguin}},\ }\href {https://doi.org/10.1103/PhysRevLett.93.076401}
  {\bibfield  {journal} {\bibinfo  {journal} {Phys. Rev. Lett.}\ }\textbf
  {\bibinfo {volume} {93}},\ \bibinfo {pages} {076401} (\bibinfo {year}
  {2004})}\BibitemShut {NoStop}\bibitem [{\citenamefont {Zgid}\ \emph {et~al.}(2012)\citenamefont {Zgid},
  \citenamefont {Gull},\ and\ \citenamefont {Chan}}]{Zgid12}\BibitemOpen
  \bibfield  {author} {\bibinfo {author} {\bibfnamefont {D.}~\bibnamefont
  {Zgid}}, \bibinfo {author} {\bibfnamefont {E.}~\bibnamefont {Gull}},\ and\
  \bibinfo {author} {\bibfnamefont {G.~K.-L.}\ \bibnamefont {Chan}},\ }\href
  {https://doi.org/10.1103/PhysRevB.86.165128} {\bibfield  {journal} {\bibinfo
  {journal} {Phys. Rev. B}\ }\textbf {\bibinfo {volume} {86}},\ \bibinfo
  {pages} {165128} (\bibinfo {year} {2012})}\BibitemShut {NoStop}\bibitem [{\citenamefont {Zhu}\ \emph {et~al.}(2019)\citenamefont {Zhu},
  \citenamefont {Jim\'enez-Hoyos}, \citenamefont {McClain}, \citenamefont
  {Berkelbach},\ and\ \citenamefont {Chan}}]{Zhu19}\BibitemOpen
  \bibfield  {author} {\bibinfo {author} {\bibfnamefont {T.}~\bibnamefont
  {Zhu}}, \bibinfo {author} {\bibfnamefont {C.~A.}\ \bibnamefont
  {Jim\'enez-Hoyos}}, \bibinfo {author} {\bibfnamefont {J.}~\bibnamefont
  {McClain}}, \bibinfo {author} {\bibfnamefont {T.~C.}\ \bibnamefont
  {Berkelbach}},\ and\ \bibinfo {author} {\bibfnamefont {G.~K.-L.}\
  \bibnamefont {Chan}},\ }\href {https://doi.org/10.1103/PhysRevB.100.115154}
  {\bibfield  {journal} {\bibinfo  {journal} {Phys. Rev. B}\ }\textbf {\bibinfo
  {volume} {100}},\ \bibinfo {pages} {115154} (\bibinfo {year}
  {2019})}\BibitemShut {NoStop}\bibitem [{\citenamefont {Shee}\ and\ \citenamefont {Zgid}(2019)}]{Shee19}\BibitemOpen
  \bibfield  {author} {\bibinfo {author} {\bibfnamefont {A.}~\bibnamefont
  {Shee}}\ and\ \bibinfo {author} {\bibfnamefont {D.}~\bibnamefont {Zgid}},\
  }\href {https://doi.org/10.1021/acs.jctc.9b00603} {\bibfield  {journal}
  {\bibinfo  {journal} {Journal of Chemical Theory and Computation}\ }\textbf
  {\bibinfo {volume} {15}},\ \bibinfo {pages} {6010} (\bibinfo {year}
  {2019})},\ \bibinfo {note} {pMID: 31518129},\ \Eprint
  {https://arxiv.org/abs/https://doi.org/10.1021/acs.jctc.9b00603}
  {https://doi.org/10.1021/acs.jctc.9b00603} \BibitemShut {NoStop}\bibitem [{\citenamefont {Shee}\ \emph {et~al.}(2021)\citenamefont {Shee},
  \citenamefont {Yeh},\ and\ \citenamefont {Zgid}}]{Shee21}\BibitemOpen
  \bibfield  {author} {\bibinfo {author} {\bibfnamefont {A.}~\bibnamefont
  {Shee}}, \bibinfo {author} {\bibfnamefont {C.-N.}\ \bibnamefont {Yeh}},\ and\
  \bibinfo {author} {\bibfnamefont {D.}~\bibnamefont {Zgid}},\ }\href@noop {}
  {\bibinfo {title} {Exploring coupled cluster green's function as a method for
  treating system and environment in green's function embedding methods}}
  (\bibinfo {year} {2021}),\ \Eprint {https://arxiv.org/abs/2107.07891}
  {arXiv:2107.07891 [physics.chem-ph]} \BibitemShut {NoStop}\bibitem [{\citenamefont {Gull}\ \emph
  {et~al.}(2011{\natexlab{a}})\citenamefont {Gull}, \citenamefont {Millis},
  \citenamefont {Lichtenstein}, \citenamefont {Rubtsov}, \citenamefont
  {Troyer},\ and\ \citenamefont {Werner}}]{Gull11_RMP}\BibitemOpen
  \bibfield  {author} {\bibinfo {author} {\bibfnamefont {E.}~\bibnamefont
  {Gull}}, \bibinfo {author} {\bibfnamefont {A.~J.}\ \bibnamefont {Millis}},
  \bibinfo {author} {\bibfnamefont {A.~I.}\ \bibnamefont {Lichtenstein}},
  \bibinfo {author} {\bibfnamefont {A.~N.}\ \bibnamefont {Rubtsov}}, \bibinfo
  {author} {\bibfnamefont {M.}~\bibnamefont {Troyer}},\ and\ \bibinfo {author}
  {\bibfnamefont {P.}~\bibnamefont {Werner}},\ }\href
  {https://doi.org/10.1103/RevModPhys.83.349} {\bibfield  {journal} {\bibinfo
  {journal} {Reviews of Modern Physics}\ }\textbf {\bibinfo {volume} {83}},\
  \bibinfo {pages} {349} (\bibinfo {year} {2011}{\natexlab{a}})}\BibitemShut
  {NoStop}\bibitem [{\citenamefont {Rubtsov}\ and\ \citenamefont
  {Lichtenstein}(2004)}]{Rubtsov04}\BibitemOpen
  \bibfield  {author} {\bibinfo {author} {\bibfnamefont {A.~N.}\ \bibnamefont
  {Rubtsov}}\ and\ \bibinfo {author} {\bibfnamefont {A.~I.}\ \bibnamefont
  {Lichtenstein}},\ }\href {https://doi.org/10.1134/1.1800216} {\bibfield
  {journal} {\bibinfo  {journal} {Journal of Experimental and Theoretical
  Physics Letters}\ }\textbf {\bibinfo {volume} {80}},\ \bibinfo {pages} {61}
  (\bibinfo {year} {2004})}\BibitemShut {NoStop}\bibitem [{\citenamefont {Rubtsov}\ \emph {et~al.}(2005)\citenamefont
  {Rubtsov}, \citenamefont {Savkin},\ and\ \citenamefont
  {Lichtenstein}}]{Rubtsov05}\BibitemOpen
  \bibfield  {author} {\bibinfo {author} {\bibfnamefont {A.~N.}\ \bibnamefont
  {Rubtsov}}, \bibinfo {author} {\bibfnamefont {V.~V.}\ \bibnamefont
  {Savkin}},\ and\ \bibinfo {author} {\bibfnamefont {A.~I.}\ \bibnamefont
  {Lichtenstein}},\ }\href {https://doi.org/10.1103/PhysRevB.72.035122}
  {\bibfield  {journal} {\bibinfo  {journal} {Physical Review B}\ }\textbf
  {\bibinfo {volume} {72}},\ \bibinfo {pages} {035122} (\bibinfo {year}
  {2005})}\BibitemShut {NoStop}\bibitem [{\citenamefont {Werner}\ \emph {et~al.}(2006)\citenamefont {Werner},
  \citenamefont {Comanac}, \citenamefont {de' Medici}, \citenamefont {Troyer},\
  and\ \citenamefont {Millis}}]{Werner06A}\BibitemOpen
  \bibfield  {author} {\bibinfo {author} {\bibfnamefont {P.}~\bibnamefont
  {Werner}}, \bibinfo {author} {\bibfnamefont {A.}~\bibnamefont {Comanac}},
  \bibinfo {author} {\bibfnamefont {L.}~\bibnamefont {de' Medici}}, \bibinfo
  {author} {\bibfnamefont {M.}~\bibnamefont {Troyer}},\ and\ \bibinfo {author}
  {\bibfnamefont {A.~J.}\ \bibnamefont {Millis}},\ }\href
  {https://doi.org/10.1103/PhysRevLett.97.076405} {\bibfield  {journal}
  {\bibinfo  {journal} {Phys. Rev. Lett.}\ }\textbf {\bibinfo {volume} {97}},\
  \bibinfo {pages} {076405} (\bibinfo {year} {2006})}\BibitemShut {NoStop}\bibitem [{\citenamefont {Werner}\ and\ \citenamefont
  {Millis}(2006)}]{Werner06B}\BibitemOpen
  \bibfield  {author} {\bibinfo {author} {\bibfnamefont {P.}~\bibnamefont
  {Werner}}\ and\ \bibinfo {author} {\bibfnamefont {A.~J.}\ \bibnamefont
  {Millis}},\ }\href {https://doi.org/10.1103/PhysRevB.74.155107} {\bibfield
  {journal} {\bibinfo  {journal} {Phys. Rev. B}\ }\textbf {\bibinfo {volume}
  {74}},\ \bibinfo {pages} {155107} (\bibinfo {year} {2006})}\BibitemShut
  {NoStop}\bibitem [{\citenamefont {Haule}(2007)}]{Haule07}\BibitemOpen
  \bibfield  {author} {\bibinfo {author} {\bibfnamefont {K.}~\bibnamefont
  {Haule}},\ }\href {https://doi.org/10.1103/PhysRevB.75.155113} {\bibfield
  {journal} {\bibinfo  {journal} {Phys. Rev. B}\ }\textbf {\bibinfo {volume}
  {75}},\ \bibinfo {pages} {155113} (\bibinfo {year} {2007})}\BibitemShut
  {NoStop}\bibitem [{\citenamefont {Gull}\ \emph {et~al.}(2008)\citenamefont {Gull},
  \citenamefont {Werner}, \citenamefont {Parcollet},\ and\ \citenamefont
  {Troyer}}]{Gull08}\BibitemOpen
  \bibfield  {author} {\bibinfo {author} {\bibfnamefont {E.}~\bibnamefont
  {Gull}}, \bibinfo {author} {\bibfnamefont {P.}~\bibnamefont {Werner}},
  \bibinfo {author} {\bibfnamefont {O.}~\bibnamefont {Parcollet}},\ and\
  \bibinfo {author} {\bibfnamefont {M.}~\bibnamefont {Troyer}},\ }\href
  {https://doi.org/10.1209/0295-5075/82/57003} {\bibfield  {journal} {\bibinfo
  {journal} {{EPL} (Europhysics Letters)}\ }\textbf {\bibinfo {volume} {82}},\
  \bibinfo {pages} {57003} (\bibinfo {year} {2008})}\BibitemShut {NoStop}\bibitem [{\citenamefont {Gull}\ \emph
  {et~al.}(2011{\natexlab{b}})\citenamefont {Gull}, \citenamefont {Staar},
  \citenamefont {Fuchs}, \citenamefont {Nukala}, \citenamefont {Summers},
  \citenamefont {Pruschke}, \citenamefont {Schulthess},\ and\ \citenamefont
  {Maier}}]{Gull11_submatrix}\BibitemOpen
  \bibfield  {author} {\bibinfo {author} {\bibfnamefont {E.}~\bibnamefont
  {Gull}}, \bibinfo {author} {\bibfnamefont {P.}~\bibnamefont {Staar}},
  \bibinfo {author} {\bibfnamefont {S.}~\bibnamefont {Fuchs}}, \bibinfo
  {author} {\bibfnamefont {P.}~\bibnamefont {Nukala}}, \bibinfo {author}
  {\bibfnamefont {M.~S.}\ \bibnamefont {Summers}}, \bibinfo {author}
  {\bibfnamefont {T.}~\bibnamefont {Pruschke}}, \bibinfo {author}
  {\bibfnamefont {T.~C.}\ \bibnamefont {Schulthess}},\ and\ \bibinfo {author}
  {\bibfnamefont {T.}~\bibnamefont {Maier}},\ }\href
  {https://doi.org/10.1103/PhysRevB.83.075122} {\bibfield  {journal} {\bibinfo
  {journal} {Phys. Rev. B}\ }\textbf {\bibinfo {volume} {83}},\ \bibinfo
  {pages} {075122} (\bibinfo {year} {2011}{\natexlab{b}})}\BibitemShut
  {NoStop}\bibitem [{\citenamefont {van Houcke}\ \emph {et~al.}(2010)\citenamefont {van
  Houcke}, \citenamefont {Kozik}, \citenamefont {Prokof'ev},\ and\
  \citenamefont {Svistunov}}]{VanHoucke2010}\BibitemOpen
  \bibfield  {author} {\bibinfo {author} {\bibfnamefont {K.}~\bibnamefont {van
  Houcke}}, \bibinfo {author} {\bibfnamefont {E.}~\bibnamefont {Kozik}},
  \bibinfo {author} {\bibfnamefont {N.}~\bibnamefont {Prokof'ev}},\ and\
  \bibinfo {author} {\bibfnamefont {B.}~\bibnamefont {Svistunov}},\ }\href
  {https://doi.org/10.1016/j.phpro.2010.09.034} {\bibfield  {journal} {\bibinfo
   {journal} {Physics Procedia}\ }\textbf {\bibinfo {volume} {6}},\ \bibinfo
  {pages} {95} (\bibinfo {year} {2010})},\ \bibinfo {note} {computer
  Simulations Studies in Condensed Matter Physics XXI}\BibitemShut {NoStop}\bibitem [{\citenamefont {Rossi}(2017)}]{Rossi2017b}\BibitemOpen
  \bibfield  {author} {\bibinfo {author} {\bibfnamefont {R.}~\bibnamefont
  {Rossi}},\ }\href {https://doi.org/10.1103/PhysRevLett.119.045701} {\bibfield
   {journal} {\bibinfo  {journal} {Physical Review Letters}\ }\textbf {\bibinfo
  {volume} {119}},\ \bibinfo {pages} {045701} (\bibinfo {year}
  {2017})}\BibitemShut {NoStop}\bibitem [{\citenamefont {Li}\ \emph {et~al.}(2020)\citenamefont {Li},
  \citenamefont {Wallerberger},\ and\ \citenamefont {Gull}}]{Li20}\BibitemOpen
  \bibfield  {author} {\bibinfo {author} {\bibfnamefont {J.}~\bibnamefont
  {Li}}, \bibinfo {author} {\bibfnamefont {M.}~\bibnamefont {Wallerberger}},\
  and\ \bibinfo {author} {\bibfnamefont {E.}~\bibnamefont {Gull}},\ }\href
  {https://doi.org/10.1103/PhysRevResearch.2.033211} {\bibfield  {journal}
  {\bibinfo  {journal} {Phys. Rev. Research}\ }\textbf {\bibinfo {volume}
  {2}},\ \bibinfo {pages} {033211} (\bibinfo {year} {2020})}\BibitemShut
  {NoStop}\bibitem [{\citenamefont {Moutenet}\ \emph {et~al.}(2018)\citenamefont
  {Moutenet}, \citenamefont {Wu},\ and\ \citenamefont
  {Ferrero}}]{Moutenet2018}\BibitemOpen
  \bibfield  {author} {\bibinfo {author} {\bibfnamefont {A.}~\bibnamefont
  {Moutenet}}, \bibinfo {author} {\bibfnamefont {W.}~\bibnamefont {Wu}},\ and\
  \bibinfo {author} {\bibfnamefont {M.}~\bibnamefont {Ferrero}},\ }\href
  {https://doi.org/10.1103/PhysRevB.97.085117} {\bibfield  {journal} {\bibinfo
  {journal} {Physical Review B}\ }\textbf {\bibinfo {volume} {97}},\ \bibinfo
  {pages} {085117} (\bibinfo {year} {2018})}\BibitemShut {NoStop}\bibitem [{\citenamefont {{\v S}imkovic}\ and\ \citenamefont
  {Kozik}(2019)}]{Simkovic2019}\BibitemOpen
  \bibfield  {author} {\bibinfo {author} {\bibfnamefont {F.}~\bibnamefont {{\v
  S}imkovic}}\ and\ \bibinfo {author} {\bibfnamefont {E.}~\bibnamefont
  {Kozik}},\ }\href {https://doi.org/10.1103/PhysRevB.100.121102} {\bibfield
  {journal} {\bibinfo  {journal} {Physical Review B}\ }\textbf {\bibinfo
  {volume} {100}},\ \bibinfo {pages} {121102} (\bibinfo {year}
  {2019})}\BibitemShut {NoStop}\bibitem [{\citenamefont {{Rossi}}(2018)}]{Rossi2018}\BibitemOpen
  \bibfield  {author} {\bibinfo {author} {\bibfnamefont {R.}~\bibnamefont
  {{Rossi}}},\ }\Eprint {https://arxiv.org/abs/1802.04743} {arXiv:1802.04743
  [cond-mat.str-el]}  (\bibinfo {year} {2018})\BibitemShut {NoStop}\bibitem [{\citenamefont {Prokof'ev}\ and\ \citenamefont
  {Svistunov}(2008)}]{Prokofev2008b}\BibitemOpen
  \bibfield  {author} {\bibinfo {author} {\bibfnamefont {N.~V.}\ \bibnamefont
  {Prokof'ev}}\ and\ \bibinfo {author} {\bibfnamefont {B.~V.}\ \bibnamefont
  {Svistunov}},\ }\href {https://doi.org/10.1103/PhysRevB.77.125101} {\bibfield
   {journal} {\bibinfo  {journal} {Physical Review B}\ }\textbf {\bibinfo
  {volume} {77}},\ \bibinfo {pages} {125101} (\bibinfo {year}
  {2008})}\BibitemShut {NoStop}\bibitem [{\citenamefont {Gull}\ \emph {et~al.}(2010)\citenamefont {Gull},
  \citenamefont {Reichman},\ and\ \citenamefont {Millis}}]{Gull10}\BibitemOpen
  \bibfield  {author} {\bibinfo {author} {\bibfnamefont {E.}~\bibnamefont
  {Gull}}, \bibinfo {author} {\bibfnamefont {D.~R.}\ \bibnamefont {Reichman}},\
  and\ \bibinfo {author} {\bibfnamefont {A.~J.}\ \bibnamefont {Millis}},\
  }\href {https://doi.org/10.1103/PhysRevB.82.075109} {\bibfield  {journal}
  {\bibinfo  {journal} {Phys. Rev. B}\ }\textbf {\bibinfo {volume} {82}},\
  \bibinfo {pages} {075109} (\bibinfo {year} {2010})}\BibitemShut {NoStop}\bibitem [{\citenamefont {Cohen}\ \emph
  {et~al.}(2014{\natexlab{a}})\citenamefont {Cohen}, \citenamefont {Reichman},
  \citenamefont {Millis},\ and\ \citenamefont {Gull}}]{Cohen2014a}\BibitemOpen
  \bibfield  {author} {\bibinfo {author} {\bibfnamefont {G.}~\bibnamefont
  {Cohen}}, \bibinfo {author} {\bibfnamefont {D.~R.}\ \bibnamefont {Reichman}},
  \bibinfo {author} {\bibfnamefont {A.~J.}\ \bibnamefont {Millis}},\ and\
  \bibinfo {author} {\bibfnamefont {E.}~\bibnamefont {Gull}},\ }\href
  {https://doi.org/10.1103/PhysRevB.89.115139} {\bibfield  {journal} {\bibinfo
  {journal} {Phys. Rev. B}\ }\textbf {\bibinfo {volume} {89}},\ \bibinfo
  {pages} {115139} (\bibinfo {year} {2014}{\natexlab{a}})}\BibitemShut
  {NoStop}\bibitem [{\citenamefont {Cohen}\ \emph
  {et~al.}(2014{\natexlab{b}})\citenamefont {Cohen}, \citenamefont {Gull},
  \citenamefont {Reichman},\ and\ \citenamefont {Millis}}]{Cohen2014b}\BibitemOpen
  \bibfield  {author} {\bibinfo {author} {\bibfnamefont {G.}~\bibnamefont
  {Cohen}}, \bibinfo {author} {\bibfnamefont {E.}~\bibnamefont {Gull}},
  \bibinfo {author} {\bibfnamefont {D.~R.}\ \bibnamefont {Reichman}},\ and\
  \bibinfo {author} {\bibfnamefont {A.~J.}\ \bibnamefont {Millis}},\ }\href
  {https://doi.org/10.1103/PhysRevLett.112.146802} {\bibfield  {journal}
  {\bibinfo  {journal} {Phys. Rev. Lett.}\ }\textbf {\bibinfo {volume} {112}},\
  \bibinfo {pages} {146802} (\bibinfo {year} {2014}{\natexlab{b}})}\BibitemShut
  {NoStop}\bibitem [{\citenamefont {Kozik}\ \emph {et~al.}(2015)\citenamefont {Kozik},
  \citenamefont {Ferrero},\ and\ \citenamefont {Georges}}]{Kozik2015}\BibitemOpen
  \bibfield  {author} {\bibinfo {author} {\bibfnamefont {E.}~\bibnamefont
  {Kozik}}, \bibinfo {author} {\bibfnamefont {M.}~\bibnamefont {Ferrero}},\
  and\ \bibinfo {author} {\bibfnamefont {A.}~\bibnamefont {Georges}},\ }\href
  {https://doi.org/10.1103/PhysRevLett.114.156402} {\bibfield  {journal}
  {\bibinfo  {journal} {Phys. Rev. Lett.}\ }\textbf {\bibinfo {volume} {114}},\
  \bibinfo {pages} {156402} (\bibinfo {year} {2015})}\BibitemShut {NoStop}\bibitem [{\citenamefont {Rossi}\ \emph {et~al.}(2018)\citenamefont {Rossi},
  \citenamefont {Ohgoe}, \citenamefont {Van~Houcke},\ and\ \citenamefont
  {Werner}}]{Rossi18B}\BibitemOpen
  \bibfield  {author} {\bibinfo {author} {\bibfnamefont {R.}~\bibnamefont
  {Rossi}}, \bibinfo {author} {\bibfnamefont {T.}~\bibnamefont {Ohgoe}},
  \bibinfo {author} {\bibfnamefont {K.}~\bibnamefont {Van~Houcke}},\ and\
  \bibinfo {author} {\bibfnamefont {F.}~\bibnamefont {Werner}},\ }\href
  {https://doi.org/10.1103/PhysRevLett.121.130405} {\bibfield  {journal}
  {\bibinfo  {journal} {Phys. Rev. Lett.}\ }\textbf {\bibinfo {volume} {121}},\
  \bibinfo {pages} {130405} (\bibinfo {year} {2018})}\BibitemShut {NoStop}\bibitem [{\citenamefont {Cohen}\ \emph {et~al.}(2015)\citenamefont {Cohen},
  \citenamefont {Gull}, \citenamefont {Reichman},\ and\ \citenamefont
  {Millis}}]{Cohen2015}\BibitemOpen
  \bibfield  {author} {\bibinfo {author} {\bibfnamefont {G.}~\bibnamefont
  {Cohen}}, \bibinfo {author} {\bibfnamefont {E.}~\bibnamefont {Gull}},
  \bibinfo {author} {\bibfnamefont {D.~R.}\ \bibnamefont {Reichman}},\ and\
  \bibinfo {author} {\bibfnamefont {A.~J.}\ \bibnamefont {Millis}},\ }\href
  {https://doi.org/10.1103/PhysRevLett.115.266802} {\bibfield  {journal}
  {\bibinfo  {journal} {Phys. Rev. Lett.}\ }\textbf {\bibinfo {volume} {115}},\
  \bibinfo {pages} {266802} (\bibinfo {year} {2015})}\BibitemShut {NoStop}\bibitem [{\citenamefont {Eidelstein}\ \emph {et~al.}(2020)\citenamefont
  {Eidelstein}, \citenamefont {Gull},\ and\ \citenamefont
  {Cohen}}]{Eidelstein2019}\BibitemOpen
  \bibfield  {author} {\bibinfo {author} {\bibfnamefont {E.}~\bibnamefont
  {Eidelstein}}, \bibinfo {author} {\bibfnamefont {E.}~\bibnamefont {Gull}},\
  and\ \bibinfo {author} {\bibfnamefont {G.}~\bibnamefont {Cohen}},\ }\href
  {https://doi.org/10.1103/PhysRevLett.124.206405} {\bibfield  {journal}
  {\bibinfo  {journal} {Phys. Rev. Lett.}\ }\textbf {\bibinfo {volume} {124}},\
  \bibinfo {pages} {206405} (\bibinfo {year} {2020})}\BibitemShut {NoStop}\bibitem [{\citenamefont {Abrikosov}\ \emph {et~al.}(1965)\citenamefont
  {Abrikosov}, \citenamefont {Gorkov},\ and\ \citenamefont
  {Dzyaloshinskii}}]{Abrikosov1965}\BibitemOpen
  \bibfield  {author} {\bibinfo {author} {\bibfnamefont {A.~A.}\ \bibnamefont
  {Abrikosov}}, \bibinfo {author} {\bibfnamefont {L.~P.}\ \bibnamefont
  {Gorkov}},\ and\ \bibinfo {author} {\bibfnamefont {I.~Y.}\ \bibnamefont
  {Dzyaloshinskii}},\ }\href@noop {} {\emph {\bibinfo {title} {Methods of
  Quantum Field Theory in Statistical Physics}}}\ (\bibinfo  {publisher}
  {Pergamon},\ \bibinfo {year} {1965})\BibitemShut {NoStop}\bibitem [{\citenamefont {Mahan}(2000)}]{Mahan2000}\BibitemOpen
  \bibfield  {author} {\bibinfo {author} {\bibfnamefont {G.~D.}\ \bibnamefont
  {Mahan}},\ }\href@noop {} {\emph {\bibinfo {title} {Many Particle Physics,
  Third Edition}}}\ (\bibinfo  {publisher} {Plenum},\ \bibinfo {address} {New
  York},\ \bibinfo {year} {2000})\BibitemShut {NoStop}\bibitem [{\citenamefont {Negele}\ and\ \citenamefont
  {Orland}(1988)}]{Negele1988}\BibitemOpen
  \bibfield  {author} {\bibinfo {author} {\bibfnamefont {J.~W.}\ \bibnamefont
  {Negele}}\ and\ \bibinfo {author} {\bibfnamefont {H.}~\bibnamefont
  {Orland}},\ }\href@noop {} {\emph {\bibinfo {title} {Quantum Many-particle
  Systems}}}\ (\bibinfo  {publisher} {Addison-Wesley},\ \bibinfo {year}
  {1988})\BibitemShut {NoStop}\bibitem [{\citenamefont {Luttinger}\ and\ \citenamefont
  {Ward}(1960)}]{Luttinger1960}\BibitemOpen
  \bibfield  {author} {\bibinfo {author} {\bibfnamefont {J.~M.}\ \bibnamefont
  {Luttinger}}\ and\ \bibinfo {author} {\bibfnamefont {J.~C.}\ \bibnamefont
  {Ward}},\ }\href {https://doi.org/10.1103/PhysRev.118.1417} {\bibfield
  {journal} {\bibinfo  {journal} {Phys. Rev.}\ }\textbf {\bibinfo {volume}
  {118}},\ \bibinfo {pages} {1417} (\bibinfo {year} {1960})}\BibitemShut
  {NoStop}\bibitem [{\citenamefont {Blankenbecler}\ \emph {et~al.}(1981)\citenamefont
  {Blankenbecler}, \citenamefont {Scalapino},\ and\ \citenamefont
  {Sugar}}]{BSS81}\BibitemOpen
  \bibfield  {author} {\bibinfo {author} {\bibfnamefont {R.}~\bibnamefont
  {Blankenbecler}}, \bibinfo {author} {\bibfnamefont {D.~J.}\ \bibnamefont
  {Scalapino}},\ and\ \bibinfo {author} {\bibfnamefont {R.~L.}\ \bibnamefont
  {Sugar}},\ }\href {https://doi.org/10.1103/PhysRevD.24.2278} {\bibfield
  {journal} {\bibinfo  {journal} {Phys. Rev. D}\ }\textbf {\bibinfo {volume}
  {24}},\ \bibinfo {pages} {2278} (\bibinfo {year} {1981})}\BibitemShut
  {NoStop}\bibitem [{\citenamefont {Boehnke}\ \emph {et~al.}(2011)\citenamefont
  {Boehnke}, \citenamefont {Hafermann}, \citenamefont {Ferrero}, \citenamefont
  {Lechermann},\ and\ \citenamefont {Parcollet}}]{boehnke2011}\BibitemOpen
  \bibfield  {author} {\bibinfo {author} {\bibfnamefont {L.}~\bibnamefont
  {Boehnke}}, \bibinfo {author} {\bibfnamefont {H.}~\bibnamefont {Hafermann}},
  \bibinfo {author} {\bibfnamefont {M.}~\bibnamefont {Ferrero}}, \bibinfo
  {author} {\bibfnamefont {F.}~\bibnamefont {Lechermann}},\ and\ \bibinfo
  {author} {\bibfnamefont {O.}~\bibnamefont {Parcollet}},\ }\href
  {https://doi.org/10.1103/PhysRevB.84.075145} {\bibfield  {journal} {\bibinfo
  {journal} {Phys. Rev. B}\ }\textbf {\bibinfo {volume} {84}},\ \bibinfo
  {pages} {075145} (\bibinfo {year} {2011})}\BibitemShut {NoStop}\bibitem [{\citenamefont {Shinaoka}\ \emph {et~al.}(2017)\citenamefont
  {Shinaoka}, \citenamefont {Otsuki}, \citenamefont {Ohzeki},\ and\
  \citenamefont {Yoshimi}}]{shinaoka2017}\BibitemOpen
  \bibfield  {author} {\bibinfo {author} {\bibfnamefont {H.}~\bibnamefont
  {Shinaoka}}, \bibinfo {author} {\bibfnamefont {J.}~\bibnamefont {Otsuki}},
  \bibinfo {author} {\bibfnamefont {M.}~\bibnamefont {Ohzeki}},\ and\ \bibinfo
  {author} {\bibfnamefont {K.}~\bibnamefont {Yoshimi}},\ }\href
  {https://doi.org/10.1103/PhysRevB.96.035147} {\bibfield  {journal} {\bibinfo
  {journal} {Phys. Rev. B}\ }\textbf {\bibinfo {volume} {96}},\ \bibinfo
  {pages} {035147} (\bibinfo {year} {2017})}\BibitemShut {NoStop}\bibitem [{\citenamefont {Gull}\ \emph {et~al.}(2018)\citenamefont {Gull},
  \citenamefont {Iskakov}, \citenamefont {Krivenko}, \citenamefont {Rusakov},\
  and\ \citenamefont {Zgid}}]{gull2018}\BibitemOpen
  \bibfield  {author} {\bibinfo {author} {\bibfnamefont {E.}~\bibnamefont
  {Gull}}, \bibinfo {author} {\bibfnamefont {S.}~\bibnamefont {Iskakov}},
  \bibinfo {author} {\bibfnamefont {I.}~\bibnamefont {Krivenko}}, \bibinfo
  {author} {\bibfnamefont {A.~A.}\ \bibnamefont {Rusakov}},\ and\ \bibinfo
  {author} {\bibfnamefont {D.}~\bibnamefont {Zgid}},\ }\href
  {https://doi.org/10.1103/PhysRevB.98.075127} {\bibfield  {journal} {\bibinfo
  {journal} {Phys. Rev. B}\ }\textbf {\bibinfo {volume} {98}},\ \bibinfo
  {pages} {075127} (\bibinfo {year} {2018})}\BibitemShut {NoStop}\bibitem [{\citenamefont {Boag}\ \emph {et~al.}(2018)\citenamefont {Boag},
  \citenamefont {Gull},\ and\ \citenamefont {Cohen}}]{Boag2018}\BibitemOpen
  \bibfield  {author} {\bibinfo {author} {\bibfnamefont {A.}~\bibnamefont
  {Boag}}, \bibinfo {author} {\bibfnamefont {E.}~\bibnamefont {Gull}},\ and\
  \bibinfo {author} {\bibfnamefont {G.}~\bibnamefont {Cohen}},\ }\href
  {https://doi.org/10.1103/PhysRevB.98.115152} {\bibfield  {journal} {\bibinfo
  {journal} {Phys. Rev. B}\ }\textbf {\bibinfo {volume} {98}},\ \bibinfo
  {pages} {115152} (\bibinfo {year} {2018})}\BibitemShut {NoStop}\bibitem [{\citenamefont {Lin}\ and\ \citenamefont {Lindsey}(2018)}]{Lin2018}\BibitemOpen
  \bibfield  {author} {\bibinfo {author} {\bibfnamefont {L.}~\bibnamefont
  {Lin}}\ and\ \bibinfo {author} {\bibfnamefont {M.}~\bibnamefont {Lindsey}},\
  }\href {https://doi.org/10.1073/pnas.1720782115} {\bibfield  {journal}
  {\bibinfo  {journal} {Proceedings of the National Academy of Sciences}\
  }\textbf {\bibinfo {volume} {115}},\ \bibinfo {pages} {2282} (\bibinfo {year}
  {2018})}\BibitemShut {NoStop}\end{thebibliography}

\end{document}